\documentclass[reprint,superscriptaddress,amsmath,amssymb,aps]{revtex4-2}
\usepackage[colorlinks=true,linkcolor=red,citecolor=blue,urlcolor=blue]{hyperref}
\usepackage{graphicx}
\usepackage{orcidlink}
\usepackage{xcolor}
\usepackage{multirow}
\begin{document}

\title{Tidal disruption or plunge: the role of neutron star internal structure in black hole–neutron star mergers}

\author{Yatharth Chaurasiya \orcidlink{0009-0002-0899-5148}}
\email{yatharth25@iiserb.ac.in}
\affiliation{Indian Institute of Science Education and Research Bhopal, Bhopal 462066, India}

\author{Ritam Mallick \orcidlink{0000-0003-2943-6388}}
\email{mallick@iiserb.ac.in}
\affiliation{Indian Institute of Science Education and Research Bhopal, Bhopal 462066, India}

\begin{abstract}
The outcome of black hole--neutron star (BHNS) mergers is studied using a general-relativistic mass-shedding condition in the Kerr spacetime, which determines the tidal-disruption radius from the mass ratio, the black hole (BH) spin, the neutron star (NS) compactness, and the ratio of central to average density, $\eta=\frac{\rho_{c}}{\bar\rho}$. Comparing this radius with the innermost stable circular orbit, we map the boundary between tidal disruption and plunge for two relativistic mean-field equations of state (EOSs) and for an ensemble of about $10^4$ agnostic EOSs spanning the current uncertainty in the NS radius. We find that the disruption threshold is set mainly by the compactness, but NSs with the same compactness and different $\eta$ can have different outcomes. Using the tidal disruption radius, we construct a remnant-mass model calibrated against 70 numerical-relativity simulations, which reproduces the remnant mass with an accuracy comparable to existing fits. Applying the disruption criterion and the remnant model to GW200105, GW200115, and the candidate BHNS event GW230529, we find that GW200105 and GW200115 are consistent with a direct plunge. GW230529 lies close to the disruption threshold and may have undergone marginal disruption, leaving only a low-mass remnant.
\end{abstract}
\maketitle

\section{Introduction}

The detection of black hole neutron star (BHNS) merger events GW200105, GW200115, and GW230529 ~\cite{GW200105_115,GW230529} by the LIGO--Virgo--KAGRA (LVK) GW detector collaboration opened a new window for studying dense matter physics through GWs \cite{LIGOwebsite,EGOwebsite,Abadie_2010,Somiya_2012}. Such systems are thought to originate either from the evolution of stellar binaries sharing a common envelope \cite{Ivanova:2012vx} or through dynamical interactions in dense stellar environments such as globular, nuclear, and young star clusters \cite{Clausen:2012zu,Ye:2019xvf,Sedda:2020wzl}. Such events can produce an electromagnetic (EM) counterpart and pave the way for multimessenger astronomy. The EM counterparts generally result from the tidal disruption of the binary's neutron star \cite{Foucart:2020ats}. The r-process nucleosynthesis generates the EM counterpart, which can range across UV/optical/infrared kilonova emission \cite{Metzger:2019zeh}, whereas the prompt short gamma-ray bursts (SGRBs)\cite{sgrb1,sgrb2} are from ultra-relativistic jets \cite{Berger:2013jza}.

The merger of black hole (BH) and neutron star (NS) usually takes one of two paths: a plunge or a disruption of NS \cite{Kyutoku:2021icp}. Whether it is a plunge or a disruption depends on the mass ratio of the BH and NS, as well as on the compactness (and thereby the EOS) \cite{matur2024signatureslowmassblack}. For a high mass ratio merger (ratio of BH to NS), the NS is seen as a point object, and the whole NS plunges into the BH as a whole and is indistinguishable from a binary BH merger. In the plunge scenario, there would be no electromagnetic counterparts, as there would be no ejecta, whereas in the disruption scenario, it is present. In the low-mass scenario, the NS is disrupted by tidal forces; its matter first forms a torus around the BH and may then form an accretion disk, a tidal tail, or ejecta. 

The analytical approach towards solving the BHNS merger is based on classical and post-Newtonian approaches. It primarily categorizes the nature of the merger: whether it is a direct plunge or a disruption \cite{Kyutoku:2021icp}. A self-consistent description of the merger and of its posible EM counterparts, however, requires numerical-relativity simulations. With this goal, several research groups have focused on general-relativistic hydrodynamical simulations of these systems \cite{matur2024signatureslowmassblack}. The first fully general relativistic simulations of BHNS mergers were
performed by Shibata and Ury\=u~\cite{ShibataUryu2006,Shibata:2006bs} and by Etienne
\textit{et al.}~\cite{Etienne2008}, and have since been carried out with several codes, including
SACRA~\cite{PhysRevD.78.064054}, SpEC~\cite{Foucart2012}, and IllinoisGRMHD~\cite{Etienne2015}.
Interest in these systems has grown considerably since the LVK Collaboration reported the first
confident detections of BHNS mergers, GW200105 and GW200115~\cite{GW200105_115}.

One of the central physical questions in BHNS mergers concerns the competition between the tidal disruption radius \cite{tidal_radius} and the innermost stable circular orbit (ISCO) radius \cite{Bardeen1972} of the BH. If the NS is tidally disrupted outside the ISCO, some matter can remain outside the BH after merger, potentially powering electromagnetic emission. If the NS crosses the ISCO before disruption, it is swallowed directly, leaving little material outside. Such processes are extremely important in illuminating the structure and properties of the NS in such binaries. Therefore, understanding the interplay among tidal forces, compactness, BH spin, and orbital dynamics is crucial for interpreting both gravitational-wave observations and their electromagnetic counterparts \cite{Kyutoku:2021icp,Foucart:2020ats}.

Subsequently, models were built to predict the remnant disk mass around the BH
and the associated GW and EM emission using numerical and analytical
calculations~\cite{Kawaguchi:2016ana,Foucart2012}.
Pannarale \textit{et al.}~\cite{Panarel} estimated the torus mass from a relativistic model of
tidal disruption, and Foucart~\cite{Foucart2012} used numerical-relativity (NR)
simulations to calibrate a fit for the remnant mass, and thus to predict whether
the NS plunges or is tidally disrupted. These predictions were subsequently
improved by recalibrating the model on a larger set of simulations with more
realistic equations of state (EOSs)~\cite{Foucart2018}, and extended to the
properties of the remnant BH and to the dynamical ejecta and disk masses~\cite{Zappa2019,Kruger2020}. In these models, tidal
disruption enters through a term motivated by the Newtonian estimate of the
tidal-disruption radius. In this work, using the tidal stretching eigenvalue
of Kerr spacetime~\cite{Marck1983,XinMummery2026}, we formulate a relativistic
mass-shedding criterion for BHNS binaries that recovers the Newtonian limit at
large separations and predicts disruption at larger separations in the
strong-field regime. Through the central-to-mean density ratio $\eta$, the
criterion captures the EOS dependence beyond the NS compactness, distinguishing
stars with similar masses and radii but different internal structures.
Incorporating this criterion into the remnant-mass model extends the
prescriptions of Refs.~\cite{Foucart2012,Foucart2018}, with an accuracy comparable to FF18 fit and a typical scatter of
$\simeq0.03\,M_{\rm NS}^{\rm b}$ for nuclear-physics-motivated EOSs, providing a framework for assessing tidal
disruption, the associated electromagnetic counterparts, and EOS constraints.

Relativistic tidal-disruption criteria for fluid stars orbiting Kerr BHs have
been obtained from equilibrium sequences of stellar
models~\cite{Fishbone1973,WigginsLai2000,Ishii2005,Ferrari2009,
Ferrari2010,Taniguchi2008}
and combined with the ISCO to estimate the remnant torus
mass~\cite{Panarel}; in these works, the disruption radius is computed
numerically for each binary configuration. Here, using the tidal stretching
eigenvalue of Kerr spacetime~\cite{Marck1983,XinMummery2026}, we formulate a
mass-shedding criterion that reduces to a single algebraic equation for the
disruption radius in terms of the mass ratio, the BH spin, and the NS
compactness and central-to-average density ratio, and we use it in place of the
phenomenological disruption term of the remnant-mass model of
FF18~\cite{Foucart2018}.

The paper is arranged as follows: In Section II, we present the formalism of our work, which provides a formulation of the mass-shedding equation and the remnant mass model within a general relativistic framework. In section III, we present our results, which encompass the disruption and plunge criterion, the remnant mass, and a parameter study of the BH and the NS. Finally, in Section IV, we summarize and draw important conclusions from our results. 

\section{Formalism}
\subsection{Mass shedding equation in general relativistic framework}
\label{subsec:mass_shedding}
In Newtonian gravity, a star orbiting a compact object of mass $M$ at a distance $r$ moves in
the potential $\Phi=-GM/r$. Because the star has a finite size, its fluid elements do not
experience the same gravitational acceleration: a fluid element displaced by $\delta x^i$
from the center of the star feels the acceleration $-\partial_i\Phi$ evaluated at
$x^i+\delta x^i$. Expanding to first order in $\delta x^i$ and subtracting the acceleration of
the center, the relative acceleration reads
\begin{equation}
\frac{d^2\delta x^i}{dt^2}=-\mathcal{E}^i{}_j\,\delta x^j,\qquad
\mathcal{E}_{ij}\equiv\frac{\partial^2\Phi}{\partial x^i\,\partial x^j},
\label{eq:tidal_newton}
\end{equation}
where $\mathcal{E}_{ij}$ is the Newtonian tidal tensor~\cite{Masi2007}. $\mathcal{E}_{ij}$ captures only the
differential gravitational pull across the star, which stretches it along the direction of
the compact object. In general relativity, the analogous role is played by the geodesic
deviation equation, and the relativistic tidal tensor leads to a relativistic mass-shedding condition.

In general relativity, the star follows a timelike geodesic of the spacetime of the compact
object, and the tidal field is described by the relative acceleration of neighboring
geodesics. In this
test-particle description the NS moves in the fixed background of the BH and its own contribution to
the spacetime is neglected. This is a good approximation for large mass ratios, but becomes
increasingly crude for $Q=M_{\rm BH}/M_{\rm NS}\lesssim2$--$3$; results at such mass ratios, and in
particular for $Q\simeq1$, should therefore be regarded as indicative only. For two nearby geodesics separated by an infinitesimal displacement vector
$\delta x^\mu$, the geodesic deviation equation reads~\cite{Weinberg1972}
\begin{equation}
\frac{D^2\delta x^\mu}{D\tau^2}=R^\mu{}_{\alpha\nu\beta}U^\alpha U^\beta\,\delta x^\nu
\equiv -C^\mu{}_\nu\,\delta x^\nu ,
\label{eq:geodesic_deviation}
\end{equation}
where $\tau$ is the proper time of the star, $U^\mu$ its four-velocity,
$R^\mu{}_{\alpha\nu\beta}$ the Riemann tensor, and $D/D\tau=U^\alpha\nabla_\alpha$ the
covariant derivative along the geodesic. Equation~(\ref{eq:geodesic_deviation}) is the
relativistic counterpart of Eq.~(\ref{eq:tidal_newton}), with the tidal tensor
$C^\mu{}_\nu\equiv-R^\mu{}_{\alpha\nu\beta}U^\alpha U^\beta$ replacing
$\partial_i\partial_j\Phi$.

The tidal accelerations experienced by the star are physically meaningful only when
$C^\mu{}_\nu$ is evaluated in the local freely falling (LFF)~\cite{Pirani2009} frame of the star, in which
$U^{\tilde\mu}=(1,0,0,0)$ and the tidal tensor reduces to
$C^{\tilde\imath}{}_{\tilde\jmath}=-R^{\tilde\imath}{}_{\tilde 0\tilde\jmath\tilde 0}$. The
traditional approach requires the construction of a parallel-propagated tetrad along the
geodesic~\cite{Marck1983}, which is in general cumbersome. Here we instead follow the method of~\cite{XinMummery2026}, which uses the zero angular momentum observer (ZAMO)
as an intermediate frame~\cite{Bardeen1972}. Any stationary, axisymmetric spacetime can be
written as
\begin{equation}
  ds^2=-e^{2\Upsilon}dt^2+e^{2\psi}(d\phi-\omega\,dt)^2+e^{2\mu_1}dr^2+e^{2\mu_2}d\theta^2,
  \label{eq:stationary_metric}
\end{equation}
where $\Upsilon$, $\psi$, $\mu_1$, $\mu_2$, and $\omega$ are functions of $r$ and $\theta$. The ZAMO,
which rotates with angular velocity $d\phi/dt=\omega$ owing to frame dragging, carries the
orthonormal tetrad
\begin{align}
  e^{(t)}&=e^{\Upsilon}\,dt, & e^{(r)}&=e^{\mu_1}\,dr,\nonumber\\
  e^{(\theta)}&=e^{\mu_2}\,d\theta, & e^{(\phi)}&=e^{\psi}\left(d\phi-\omega\,dt\right),
  \label{eq:zamo_tetrad}
\end{align}
in which the metric is locally Minkowskian, $g_{(a)(b)}=\eta_{(a)(b)}$. The procedure then
consists of three steps: (i) the Riemann tensor and the four-velocity of the star are
projected onto the ZAMO tetrad; (ii) the Riemann tensor is transformed to the LFF frame of
the star by a Lorentz boost with the three-velocity of the star relative to the ZAMO,
$\beta^{(i)}=U^{(i)}/U^{(t)}$; and (iii) the local tidal tensor is constructed as
$C^{\tilde\imath}{}_{\tilde\jmath}=-R^{\tilde\imath}{}_{\tilde 0\tilde\jmath\tilde 0}$.
Since only standard frame transformations are involved, this yields the tidal tensor in
closed form for any point of the orbit, without the need of a parallel-propagated
tetrad~\cite{XinMummery2026}.

The tidal tensor in the LFF frame always has a vanishing eigenvalue along the time
direction, since $C^{\tilde 0}{}_{\tilde\nu}=0$, and its three spatial eigenvalues
$\lambda_k$ represent the physical tidal accelerations per unit length along the principal
directions, in analogy with the Newtonian case. A negative eigenvalue corresponds to
stretching and a positive one to compression~\cite{XinMummery2026}.

Tidal disruption of the NS sets in when the tidal acceleration along the
direction of maximal stretching, evaluated at the stellar surface, exceeds the
self-gravitational acceleration that binds the outermost layers to the star,
so that mass is stripped from the surface (mass shedding).
In the LFF frame, a fluid element at the stellar surface, at a distance $R_\star$ from the
center along the stretching direction, experiences the tidal acceleration
$|\lambda_{\min}|\,R_\star$, where $\lambda_{\min}<0$ is the most negative eigenvalue of
$C^{\tilde\imath}{}_{\tilde\jmath}$. Mass shedding therefore sets in when
\begin{equation}
|\lambda_{\min}|\,R_\star\;\geq\;\eta\,\frac{M_\star}{R_\star^{2}},
\label{eq:mass_shedding}
\end{equation}
where $M_\star$ and $R_\star$ are the mass and radius of the star, and the
right-hand side represents its self-gravitational acceleration at the surface.
Following Ref.~\cite{XinMummery2026}, the dimensionless parameter $\eta$ can be
calibrated with hydrodynamical simulations of tidal
disruption~\cite{Guillochon_2013}, and it is expected to be well approximated by
the ratio of the central to the average density of the star,
\begin{equation}
  \eta\simeq\frac{\rho_c}{\bar\rho},\qquad\bar\rho=\frac{3M_\star}{4\pi R_\star^3},
  \label{eq:eta_rho}
\end{equation}
which we compute for each EoS by solving the TOV
equations~\cite{TOV1,TOV2}. The commonly adopted choice
$\eta=1$~\cite{29,32} is a simplification. In hydrodynamical simulations of
stellar tidal disruption, the tidal strength required to disrupt a star increases with its central
concentration, and factors of this type, $\eta>1$, were introduced to describe this
dependence~\cite{Guillochon_2013,33}. We note that these calibrations refer to the disruption of
the stellar core, whereas Eq.~\eqref{eq:mass_shedding} describes the onset of mass shedding from the
surface. The surface gravity $M_\star/R_\star^2$ itself does not depend on the density profile; in
our criterion $\eta$ is therefore an effective parameter that encodes how the response of the star to
the tidal field depends on its internal structure, and it is a single-parameter description of that
structure. Stars with a dense core and an extended, weakly bound envelope, such as $\Gamma=2$
polytropes, may shed mass more easily than a large $\eta$ implies; we return to this point in
Sec.~\ref{sec:rem_mass_results}.

In units $G=c=1$, masses and radii have dimensions of length, the tidal eigenvalues have
dimensions of $[L]^{-2}$, and both sides of Eq.~(\ref{eq:mass_shedding}) have
dimensions of $[L]^{-1}$, i.e., of an acceleration. In general, we need to determine the negative eigenvalues for the chosen metric.

We consider a rotating BH of mass $M$ and spin parameter $a=\chi M$, described by the Kerr
metric~\cite{Chandrasekhar1983} in Boyer--Lindquist coordinates ($G=c=1$),
\begin{align}
ds^2=&-\left(1-\frac{2Mr}{\Sigma}\right)dt^2
-\frac{4Mar\sin^2\theta}{\Sigma}\,dt\,d\phi
+\frac{\Sigma}{\Delta}\,dr^2 \nonumber\\
&+\Sigma\,d\theta^2
+\left(r^2+a^2+\frac{2Ma^2r\sin^2\theta}{\Sigma}\right)\sin^2\theta\,d\phi^2 ,
\label{eq:Kerr_metric}
\end{align}
where $\Sigma=r^2+a^2\cos^2\theta$ and $\Delta=r^2-2Mr+a^2$.

Geodesic motion in the Kerr spacetime admits three constants of motion per unit rest mass:
the energy $\epsilon=-u_t$, the axial angular momentum $l_z=u_\phi$, and the Carter
constant~\cite{Carter1968}
\begin{equation}
\mathcal{Q}=u_\theta^2+\cos^2\theta\left[a^2\left(1-\epsilon^2\right)
+\frac{l_z^2}{\sin^2\theta}\right],
\label{eq:Carter}
\end{equation}
which is conveniently used in the non-negative form
$k=\mathcal{Q}+(l_z-a\epsilon)^2$. The radial motion is governed by
\begin{equation}
\Sigma^2\left(\frac{dr}{d\tau}\right)^2=R(r),
\label{eq:Kerr_radial_eom}
\end{equation}
with the radial potential
\begin{equation}
R(r)=\left[\epsilon(r^2+a^2)-al_z\right]^2-\Delta\left(r^2+k\right).
\label{eq:Kerr_R}
\end{equation}
We consider circular orbits on the equatorial plane, $\theta=\pi/2$. Since such orbits have
$u_\theta=0$ and $\cos\theta=0$, both terms in Eq.~(\ref{eq:Carter}) vanish, so that
$\mathcal{Q}=0$ and $k=(l_z-a\epsilon)^2$. Circular orbits further require $R(r)=R'(r)=0$,
which yields~\cite{Chandrasekhar1983}
\begin{align}
\epsilon&=\frac{r^{3/2}-2Mr^{1/2}+aM^{1/2}}
{r^{3/4}\left(r^{3/2}-3Mr^{1/2}+2aM^{1/2}\right)^{1/2}},
\label{eq:Kerr_eps}\\
l_z&=\frac{M^{1/2}\left(r^2-2aM^{1/2}r^{1/2}+a^2\right)}
{r^{3/4}\left(r^{3/2}-3Mr^{1/2}+2aM^{1/2}\right)^{1/2}},
\label{eq:Kerr_lz}
\end{align}
for prograde orbits, with retrograde orbits obtained by $a\to-a$. Combining
Eqs.~(\ref{eq:Kerr_eps}) and~(\ref{eq:Kerr_lz}), the Carter constant of the orbit becomes
\begin{equation}
k=(l_z-a\epsilon)^2=\frac{r^{3/2}\left(\sqrt{Mr}-a\right)^2}
{r^{3/2}-3Mr^{1/2}+2aM^{1/2}} .
\label{eq:Kerr_k}
\end{equation}
The negative eigenvalue of the local tidal tensor in the Kerr spacetime, obtained with the
ZAMO algorithm in ~\cite{XinMummery2026}, which is responsible for the stretching of the
star, on the equatorial plane is
\begin{equation}
\lambda_{\min}=-\frac{M}{r^3}\left(2+\frac{3k}{r^2}\right).
\label{eq:Kerr_lambda_general}
\end{equation}
Substituting the Carter constant of circular equatorial orbits, Eq.~(\ref{eq:Kerr_k}), we
obtain
\begin{align}
\lambda_{\min}&=-\frac{M}{r^3}\left[2+3\,\frac{Mr^{1/2}+a^2r^{-1/2}-2aM^{1/2}}
{r^{3/2}-3Mr^{1/2}+2aM^{1/2}}\right] \nonumber\\
&=-\frac{M}{r^3}\,\frac{2r^2-3Mr-2a\sqrt{Mr}+3a^2}{r^2-3Mr+2a\sqrt{Mr}},
\label{eq:Kerr_lambda}
\end{align}
Substituting Eq.~(\ref{eq:Kerr_lambda}) into the mass-shedding criterion,
Eq.~(\ref{eq:mass_shedding}), the onset of tidal disruption is given by
\begin{equation}
\frac{M}{r^3}\left[2+3\,\frac{Mr^{1/2}+a^2r^{-1/2}-2aM^{1/2}}
{r^{3/2}-3Mr^{1/2}+2aM^{1/2}}\right]R_\star=\eta\,\frac{M_\star}{R_\star^2}.
\label{eq:TD_balance}
\end{equation}
Multiplying by $r^3/(M_\star R_\star)$ and introducing the mass ratio $Q\equiv M/M_\star$,
the critical threshold for mass shedding reads
\begin{equation}
-\eta\,\frac{r^3}{R_\star^3}+2Q+3Q\,
\frac{QM_\star\,r^{1/2}+a^2r^{-1/2}-2a\left(QM_\star\right)^{1/2}}
{r^{3/2}-3QM_\star\,r^{1/2}+2a\left(QM_\star\right)^{1/2}}=0 .
\label{eq:TD_threshold}
\end{equation}
If the companion of the BH is a NS, Eq.~(\ref{eq:TD_threshold}) applies with $M_\star$ and
$R_\star$ replaced by the NS mass $M_{\rm NS}$ and radius $R_{\rm NS}$, respectively
($Q=M_{\rm BH}/M_{\rm NS}$):
\begin{equation}
-\eta\,\frac{r^3}{R_{\rm NS}^3}+2Q+3Q\,\frac{\mathcal{N}(r)}{\mathcal{D}(r)}=0 ,
\label{eq:TD_threshold_NS}
\end{equation}
where
\begin{align}
\mathcal{N}(r)&=QM_{\rm NS}\,r^{1/2}+a^2r^{-1/2}-2a\left(QM_{\rm NS}\right)^{1/2},
\nonumber\\
\mathcal{D}(r)&=r^{3/2}-3QM_{\rm NS}\,r^{1/2}+2a\left(QM_{\rm NS}\right)^{1/2}.
\label{eq:ND_def}
\end{align}
Equation~(\ref{eq:TD_threshold_NS}) is the mass-shedding condition, and its largest real
root, $r=r_{\rm TD}$, defines the threshold tidal disruption radius in the Kerr spacetime.
For the NS to be tidally disrupted before plunging into the BH, the disruption must occur
outside the ISCO, i.e., $r_{\rm TD}\geq r_{\rm ISCO}$. The ISCO radius of a Kerr BH is the
root of~\cite{Chandrasekhar1983}
\begin{equation}
r^2-6Mr\pm8a\sqrt{Mr}-3a^2=0 ,
\label{eq:ISCO}
\end{equation}
where the upper (lower) sign corresponds to prograde (retrograde) orbits; for $a=0$ it
reduces to the Schwarzschild value $r_{\rm ISCO}=6M$.

To examine the Newtonian limit of the mass-shedding condition,
Eq.~(\ref{eq:TD_threshold_NS}), we divide the numerator and denominator of
Eq.~(\ref{eq:ND_def}) by $r^{3/2}$, which gives
\begin{equation}
\frac{\mathcal{N}(r)}{\mathcal{D}(r)}=
\frac{\dfrac{QM_{\rm NS}}{r}+\dfrac{a^2}{r^2}-\dfrac{2a\left(QM_{\rm NS}\right)^{1/2}}{r^{3/2}}}
{1-\dfrac{3QM_{\rm NS}}{r}+\dfrac{2a\left(QM_{\rm NS}\right)^{1/2}}{r^{3/2}}} .
\label{eq:ND_expanded}
\end{equation}
We introduce the small parameter
\begin{equation}
\zeta\equiv\frac{M_{\rm BH}}{r}=\frac{QM_{\rm NS}}{r},
\label{eq:epsilon}
\end{equation}
which measures the strength of the gravitational field of the BH at the position of the
NS, and write the spin as $a=\chi M_{\rm BH}$, with $|\chi|\leq1$. The terms in
Eq.~(\ref{eq:ND_expanded}) then become $QM_{\rm NS}/r=\zeta$,
$a^2/r^2=\chi^2\zeta^2$, and $2a(QM_{\rm NS})^{1/2}/r^{3/2}=2\chi\,\zeta^{3/2}$.
Since $|\chi|\leq1$, the spin-dependent terms are bounded by
$\chi^2\zeta^2\leq\zeta^2$ and $|2\chi\,\zeta^{3/2}|\leq2\zeta^{3/2}$,
independently of the BH spin. Expanding for $\zeta\ll1$, we obtain
\begin{align}
\frac{\mathcal{N}}{\mathcal{D}}
&=\zeta\,\frac{\left(1-\chi\sqrt{\zeta}\right)^2}
{1-3\zeta+2\chi\,\zeta^{3/2}} \nonumber\\
&=\zeta-2\chi\,\zeta^{3/2}+\left(3+\chi^2\right)\zeta^{2}
+\mathcal{O}\!\left(\zeta^{5/2}\right).
\label{eq:ND_series}
\end{align}
The relativistic term is therefore of order $\zeta$, while the BH spin enters only at
order $\zeta^{3/2}$. In the limit of large separation, $r\to\infty$, we have
$\zeta\to0$, and the whole relativistic term vanishes. Note that the nonspinning limit
$\chi\to0$ does not lead to the Newtonian result, since it removes only the spin-dependent
terms in Eq.~(\ref{eq:ND_series}) and recovers the Schwarzschild spacetime.

In this limit, $\mathcal{N}/\mathcal{D}\to0$ and the mass-shedding condition reduces to
$-\eta\,r^3/R_{\rm NS}^3+2Q=0$, whose solution is the Newtonian tidal radius,
\begin{equation}
r_{\rm TD}^{\rm N}=\left(\frac{2Q}{\eta}\right)^{1/3}R_{\rm NS}
=\left(\frac{2M_{\rm BH}}{\eta\,M_{\rm NS}}\right)^{1/3}R_{\rm NS} .
\label{eq:rTD_newton}
\end{equation}
Retaining the leading term of Eq.~(\ref{eq:ND_series}),
$\mathcal{N}/\mathcal{D}\simeq\zeta$, yields the first relativistic correction,
\begin{equation}
r_{\rm TD}\simeq r_{\rm TD}^{\rm N}\left(1+\frac{\zeta_{\rm N}}{2}\right),
\qquad \zeta_{\rm N}=\frac{M_{\rm BH}}{r_{\rm TD}^{\rm N}},
\label{eq:rTD_1PN}
\end{equation}
which shows that relativistic tidal forces shift the onset of mass shedding to larger
separations than predicted by Newtonian gravity.

The validity of the Newtonian limit for BHNS binaries can be assessed self-consistently by
evaluating the expansion parameter at the Newtonian tidal radius. In physical units,
$\zeta=GM_{\rm BH}/(c^2r)=|\Phi_{\rm N}|/c^2$ is the Newtonian gravitational potential
of the BH at the position of the NS in units of $c^2$, and for a circular orbit it also
measures the orbital velocity, $v^2/c^2\simeq\zeta$. It therefore quantifies both the
strength of the gravitational field and the speed of the orbital motion. As shown by
Eq.~(\ref{eq:ND_series}), all relativistic contributions to the mass-shedding condition
scale as powers of $\zeta$, the leading one representing a fractional correction of
$\simeq\tfrac{3}{2}\zeta$ to the Newtonian term. The Newtonian description is thus
accurate only in the weak-field, slow-motion regime, $\zeta\ll1$, and this condition
must hold at the separation where the NS is predicted to be disrupted. Using
Eq.~(\ref{eq:rTD_newton}), the expansion parameter at the Newtonian tidal radius is
\begin{equation}
\zeta_{\rm N}=\frac{M_{\rm BH}}{r_{\rm TD}^{\rm N}}
=\frac{QC}{(2Q/\eta)^{1/3}}
=C\,Q^{2/3}\left(\frac{\eta}{2}\right)^{1/3},
\label{eq:epsN}
\end{equation}
where $C=M_{\rm NS}/R_{\rm NS}$ is the compactness of the NS. For typical BHNS parameters,
$Q=3$--$7$, $C\simeq0.15$--$0.18$, and $\eta\simeq2$, the Newtonian tidal radius lies at only
$r_{\rm TD}^{\rm N}\simeq(1.4$--$1.9)\,R_{\rm NS}$, and Eq.~(\ref{eq:epsN}) gives
$\zeta_{\rm N}\simeq0.3$--$0.7$. Since $\zeta_{\rm N}$ is not small, the
relativistic terms are comparable to the Newtonian one, and the Newtonian estimate fails
precisely where it predicts disruption. For instance, for $Q=3$ and a NS with $C=0.15$ and $\eta=2.1$ orbiting a BH with
$\chi=0.8$, the full mass-shedding condition, Eq.~\eqref{eq:TD_threshold_NS}, yields
$r_{\rm TD}\simeq1.57\,R_{\rm NS}$, about $11\%$ larger than the Newtonian estimate,
$r_{\rm TD}^{\rm N}\simeq1.42\,R_{\rm NS}$, and outside the ISCO, $r_{\rm ISCO}\simeq1.31\,R_{\rm NS}$.
The relativistic enhancement grows as the spin decreases, but for the same binary with $\chi=0$ the
root, $r_{\rm TD}\simeq1.87\,R_{\rm NS}$, lies inside the ISCO ($r_{\rm ISCO}=2.7\,R_{\rm NS}$), where
circular orbits are unstable and the NS plunges. When $r_{\rm TD}<r_{\rm ISCO}$, the value of
$r_{\rm TD}$ therefore has no direct meaning as a disruption radius; it enters the disruption criterion
only through the sign of $r_{\rm TD}-r_{\rm ISCO}$, and the remnant-mass model of
Sec.~\ref{sec:Mrem_model} only as a smooth fitting variable. A relativistic treatment of the tidal
field is therefore essential for BHNS mergers. Throughout the remainder of this paper, $r_{\rm TD}$
denotes the largest real root of the mass-shedding condition, Eq.~\eqref{eq:TD_threshold_NS}, and
$r_{\rm ISCO}$ the ISCO radius of the BH, given by the root of Eq.~\eqref{eq:ISCO}, both evaluated for
the same set of binary parameters $(Q,\chi_{\rm BH},C,\eta)$.
\subsection{Remnant mass model}
\label{sec:Mrem_model}

The mass remaining outside the BH at late times in a BHNS merger is determined by the
competition between the tidal disruption radius and the ISCO. Part of the disrupted
material forms an accretion disk, part is ejected into a tidal tail and falls back onto the
remnant over longer timescales, and a fraction may become gravitationally unbound. Fitting
formulas for the total remnant mass, calibrated to NR simulations, were proposed in
Refs.~\cite{Foucart2012,Foucart2018}. In this work we adopt the nonlinear functional form of
Ref.~\cite{Foucart2018} (hereafter FF18),
\begin{align}
\hat{M}^{\rm rem}_{\rm model}&\equiv\frac{M^{\rm rem}_{\rm model}}{M^{\rm b}_{\rm NS}}
\nonumber\\
&=\biggl[\max\biggl(\alpha\,\frac{r_{\rm TD}}{R_{\rm NS}}
-\beta\,\frac{r_{\rm ISCO}}{R_{\rm NS}}+\gamma,\,0\biggr)\biggr]^{\delta},
\label{eq:Mrem_general}
\end{align}
where $M^{\rm rem}_{\rm model}$ is the remnant mass predicted by the model,
$M^{\rm b}_{\rm NS}$ is the baryonic mass of the NS, $r_{\rm TD}$ is the tidal disruption
radius, i.e., the separation at which mass shedding sets in, and $r_{\rm ISCO}$ is the ISCO
radius of the BH. The coefficients $\alpha$, $\beta$, $\gamma$, and $\delta$ are free
parameters determined by a least-squares fit to the results of NR simulations.

In FF18, the two radii are estimated in terms of the compactness of the NS,
$C=M_{\rm NS}/R_{\rm NS}$, as
\begin{equation}
\hat{M}^{\rm rem}_{\rm model}=\left[\max\left(\alpha\frac{1-2C}{\nu^{1/3}}
-\beta\hat{r}_{\rm ISCO}\frac{C}{\nu}+\gamma,0\right)\right]^{\delta},
\label{eq:Mrem_FF18}
\end{equation}
where $\nu=Q/(1+Q)^2$ is the symmetric mass ratio and
$\hat{r}_{\rm ISCO}\equiv r_{\rm ISCO}/M_{\rm BH}$ is the ISCO radius in units of the BH
mass. The first term is a Newtonian-like estimate of
$r_{\rm TD}/R_{\rm NS}$, with the factor $1-2C$ accounting approximately for relativistic
effects. Alternatively, FF18~\cite{Foucart2018}
expressed the model in terms of the compactness-like parameter $\rho=(15\Lambda)^{-1/5}$,
defined from the dimensionless tidal deformability $\Lambda$ of the NS, by replacing $C$ with
$\rho$ in Eq.~(\ref{eq:Mrem_FF18}). Since $\rho$ differs from $C$ by only a few percent, the
two forms are nearly equivalent, but the latter allows the remnant mass to be expressed
directly in terms of $\Lambda$, which is measured in GW observations~\cite{GW170817,GW170817_EOS,GW170817_properties}.

In our remnant mass model, the tidal disruption radius $r_{\rm TD}$ is obtained from the
relativistic mass-shedding condition, Eq.~(\ref{eq:TD_threshold_NS}), while the ISCO term
is taken as $\hat{r}_{\rm ISCO}\,\rho/\nu$, following the tidal-deformability form of FF18.
Normalizing the separation by the NS radius, $\xi\equiv r/R_{\rm NS}$, and using
$M_{\rm BH}=QC\,R_{\rm NS}$ and $a=\chi_{\rm BH}M_{\rm BH}$, the mass-shedding condition
takes the dimensionless form
\begin{align}
&-\eta\,\xi^3+2Q \nonumber\\
&+3Q\,\frac{QC\,\xi^{1/2}+(\chi_{\rm BH}QC)^2\xi^{-1/2}-2\chi_{\rm BH}(QC)^{3/2}}
{\xi^{3/2}-3QC\,\xi^{1/2}+2\chi_{\rm BH}(QC)^{3/2}}=0 ,
\label{eq:mass_shedding_normalized}
\end{align}
where $C=M_{\rm NS}/R_{\rm NS}$ is the compactness of the NS. The normalized tidal
disruption radius, $\xi_{\rm TD}\equiv r_{\rm TD}/R_{\rm NS}$, is defined as the largest real
root of Eq.~(\ref{eq:mass_shedding_normalized}). The remnant mass model then reads
\begin{equation}
\hat{M}^{\rm rem}_{\rm model}=\left[\max\left(\alpha\,\xi_{\rm TD}
-\beta\,\hat{r}_{\rm ISCO}\,\frac{\rho}{\nu}+\gamma,\,0\right)\right]^{\delta},
\label{eq:Mrem_model}
\end{equation}
Equation~(\ref{eq:Mrem_model}) differs from FF18 in that the
Newtonian-like estimate of the tidal radius is replaced by the solution of the relativistic
mass-shedding condition, which includes the internal structure of the NS through $\eta$. We calibrate the remnant mass model of Eq.~(\ref{eq:Mrem_model}) against the results of NR
simulations of BHNS mergers performed by several groups, as described in
Sec.~\ref{sec:NR_data}, and use the calibrated model to predict the remnant mass for the
binary configurations considered in this work.
\subsection{Comparison with previous work}
\label{sec:prev_work}

Our mass-shedding criterion is closely related to earlier relativistic
treatments of the tidal problem. Wiggins and Lai~\cite{WigginsLai2000} employed
the same Kerr tidal tensor~\cite{Marck1983} to construct equilibrium sequences
of compressible ellipsoids on circular equatorial orbits, while Ishii
\textit{et al.}~\cite{Ishii2005} computed Roche limits of polytropic stars in
Fermi normal coordinates, including tidal terms up to fourth order in
$R_{\rm NS}/r$. Both studies found that relativistic effects move the onset of
disruption outward relative to the Newtonian estimate, and that prograde BH
spin reduces this shift.

The main difference from these works is that our criterion is a single
algebraic equation in $(Q,\chi,C,\eta)$ rather than a numerical equilibrium
calculation. The NS structure enters through
$\eta=\rho_{c}/\bar\rho$, which we compute from TOV solutions of
realistic EOSs. This makes it possible to evaluate the disruption radius over
large EOS ensembles and to calibrate it directly against numerical-relativity
simulations.

Our remnant model retains the functional form of the FF18
model~\cite{Foucart2018} but differs in how tidal disruption enters. In FF18,
the disruption term, $\alpha\,(1-2C)/\nu^{1/3}$, is a phenomenological
scaling motivated by the Newtonian tidal radius,
$r_{\rm TD}\propto Q^{1/3}R_{\rm NS}$, with the factor $(1-2C)$ accounting
approximately for the compactness of the NS; it is independent of the BH spin,
which enters only through $\hat{r}_{\rm ISCO}$. In our model, this term is
replaced by $\alpha\,\xi_{\rm TD}$, where $\xi_{\rm TD}$ is the disruption
radius obtained from the relativistic mass-shedding condition,
Eq.~(\ref{eq:mass_shedding_normalized}). The disruption term thus depends explicitly
on the BH spin through the Kerr tidal field and on the internal structure of
the NS through $\eta$, while the ISCO term,
$-\beta\,\hat{r}_{\rm ISCO}\,\rho/\nu$, is kept as in FF18. Both terms are
therefore physically motivated, and the four coefficients
$(\alpha,\beta,\gamma,\delta)$ are recalibrated on the same NR data set.

\subsection{Equations of state and numerical-relativity data}
\label{sec:NR_data}


The EOS dependence of the outcome, whether plunge or tidal disruption, is investigated using both mean-field and agnostic EOS models. For the mean-field approach, we consider two representative EOSs, DDB and RMFNL~\cite{sanson_hyperonic_2026}. The DDB EOS employs density-dependent meson–nucleon couplings, as introduced in \cite{Malik:2022}, whereas the RMFNL EOS assumes constant meson–nucleon couplings and incorporates nonlinear meson terms in the Lagrangian density~\cite{sanson_hyperonic_2026}. For the agnostic approach, we employ the speed-of-sound interpolation framework introduced in \cite{Annala:2019puf}, considering both monotonic and non-monotonic classes of EOSs as described in \cite{Verma:2025dez}. The parameters defining the agnostic EOS ensemble are obtained from the publicly available repository of \cite{asim_kumar_saha_2026}.

To determine the parameters $\alpha$, $\beta$, $\gamma$, and $\delta$ of the remnant mass
model, Eq.~(\ref{eq:Mrem_model}), we use the results of NR simulations of BHNS mergers performed by several groups
with three independent evolution codes, SACRA, SpEC, and the UIUC
code~\cite{NR1,NR2,NR3,NR4,NR5,NR6,NR7,NR8,NR9}, as compiled in
Ref.~\cite{Foucart2018}. Of the 75 simulations listed there (see the
Supplemental Material of Ref.~\cite{Foucart2018} for details), we retain 70 in
our analysis. We exclude the two
configurations with $Q=1$ and $Q=1.2$, and the three configurations whose results combine two
different codes (SpEC/SACRA, UIUC/SpEC, and UIUC/SACRA). Of the remaining
simulations, 49 were performed with SACRA, 19 with SpEC, and 2 with the UIUC
code. Each simulation is characterized by the mass ratio
$Q=M_{\rm BH}/M_{\rm NS}$, the dimensionless spin of the BH,
$\chi_{\rm BH}=J_{\rm BH}/M_{\rm BH}^{2}$, aligned (or anti-aligned) with the
orbital angular momentum, and the compactness of the NS,
$C=M_{\rm NS}/R_{\rm NS}$. The retained set spans $3\le Q\le7$,
$-0.5\le\chi_{\rm BH}\le0.97$, and $0.130\le C\le0.182$; of these simulations,
52 employ piecewise-polytropic EOSs, 10 employ tabulated nuclear-physics EOSs,
and 8 employ $\Gamma=2$ polytropes.
\section{Results}
\subsection{Tidal disruption or Plunge}
\label{sec:TD_or_plunge}
\begin{figure*}
    \centering
    \includegraphics[width=0.48\textwidth]{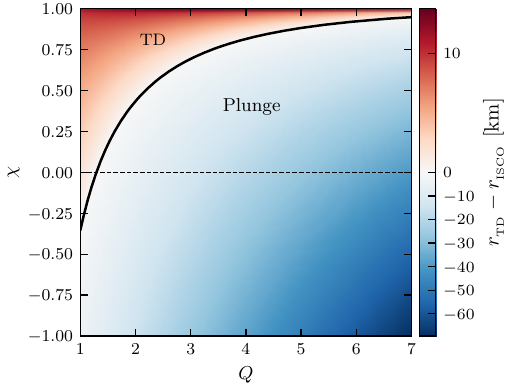}
    \includegraphics[width=0.48\textwidth]{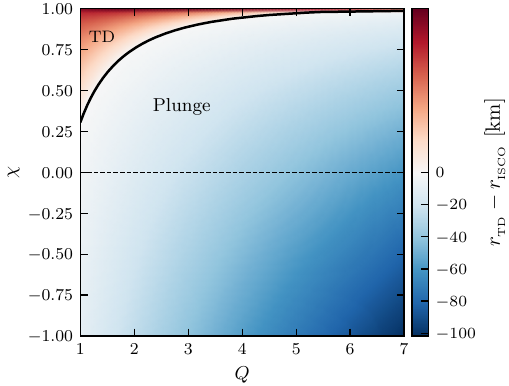}
\caption{Tidal disruption (TD) and plunge criteria for the DDB EOS. Left: neutron star (NS) with mass $1.4\,M_\odot$ ($R_{\rm NS} = 13.14\,\mathrm{km}$, $\eta = 2.13$). Right: NS with mass $2.0\,M_\odot$ ($R_{\rm NS} = 13.09\,\mathrm{km}$, $\eta = 2.15$). The black solid line is the critical point where tidal radius is equal to the ISCO radius, and the colour map shows the difference
    $r_{\rm TD}-r_{\rm ISCO}$. The dashed line marks a nonspinning BH, $\chi=0$. Binaries above the
    solid line are tidally disrupted.}
    \label{fig:DDB_secI}
\end{figure*}

We study the tidal disruption (TD) and plunge criteria for a fixed NS mass using the DDB EOS, while varying the mass ratio $Q=M_{\rm BH}/M_{\rm NS}$ and the BH spin $\chi$, using the mass-shedding condition, Eq.(~\ref{eq:TD_threshold_NS}), as shown in Fig.~\ref{fig:DDB_secI}. As expected, more compact NSs require a higher prograde BH spin to be tidally disrupted: a more compact NS has a smaller tidal disruption radius, so the ISCO must be correspondingly smaller for disruption to occur outside it, which is achieved for higher prograde BH spins. The same analysis has been done for the RMFNL, monotonic, and nonmonotonic EOSs. Table~\ref{tab:TD_summary} summarizes the NS properties and the minimum BH spin required
for tidal disruption for each EOS. For a given EOS, increasing the NS mass from
$1.4\,M_\odot$ to $2.0\,M_\odot$ increases the compactness of the star, and the minimum
spin required for disruption at $Q=1$ shifts from retrograde to prograde values (as noted in
Sec.~\ref{subsec:mass_shedding}, values at $Q\simeq1$ are only indicative); for the DDB EOS, for instance, it increases from
$\chi_{\rm BH}=-0.35$ to $\chi_{\rm BH}=0.30$. This behavior follows from the competition
between the tidal disruption radius and the ISCO. A more massive, more compact NS is more
strongly self-bound, so that the tidal acceleration required to strip its surface layers
is reached only closer to the BH, i.e., $r_{\rm TD}$ decreases. Since the ISCO radius
depends only on the BH mass and spin, the condition $r_{\rm TD}\geq r_{\rm ISCO}$ can then
be satisfied only if the ISCO is moved inward, which requires a larger prograde BH spin.
\begin{table*}[!t]
  \caption{Neutron star properties and the minimum aligned BH spin required for tidal disruption for
    each EOS. $\chi_{\rm BH}$ is the minimum aligned BH spin for which tidal disruption occurs at mass
    ratio $Q=1$ and $Q=7$. $M_{\rm BH}^{\chi=0}$ is the maximum BH mass that allows disruption for a
    nonspinning BH; a dash indicates that a nonspinning BH cannot disrupt the NS for any $Q\geq1$.
    Values at $Q\simeq1$ lie outside the test-particle regime and are only indicative.}
  \label{tab:eos_spin}
  \begin{ruledtabular}
  \begin{tabular}{lcccccccc}
    EOS & $M_{\rm NS}\,[M_\odot]$ & $R_{\rm NS}\,[\mathrm{km}]$ & $C$ & $\eta$ &
    $\chi_{\rm BH}\,(Q=1)$ & $\chi_{\rm BH}\,(Q=7)$ & $M_{\rm BH}^{\chi=0}\,[M_\odot]$ \\
    \hline
    DDB           & 1.4 & 13.14 & 0.157 & 2.13 & $-0.35$ & 0.95 & 1.81 \\
                  & 2.0 & 13.09 & 0.225 & 2.15 & $0.30$  & 0.99 & ---  \\
    \hline
    RMFNL         & 1.4 & 12.78 & 0.162 & 2.04 & $-0.32$ & 0.95 & 1.77 \\
                  & 2.0 & 12.79 & 0.230 & 2.09 & $0.33$  & 0.99 & ---  \\
    \hline
    Monotonic     & 1.4 & 11.25 & 0.184 & 2.08 & $-0.04$ & 0.97 & 1.45 \\
                  & 2.0 & 11.19 & 0.264 & 1.89 & $0.45$  & 0.99 & ---  \\
    \hline
    Non-monotonic & 1.4 & 11.20 & 0.184 & 1.93 & $-0.08$ & 0.97 & 1.50 \\
                  & 2.0 & 11.35 & 0.260 & 1.81 & $0.42$  & 0.99 & ---  \\
  \end{tabular}
  \end{ruledtabular}
\end{table*}
\begin{table*}
\caption{\label{tab:gw_events}Source properties of the BHNS events used in this work~\cite{GW200105_115,GW230529}: median values with
90\% credible intervals of the BH mass $M_{\rm BH}$, NS mass $M_{\rm NS}$ and aligned BH spin
$\chi=\chi_{1z}$, for the high-spin ($\chi_{2}<0.99$) and low-spin ($\chi_{2}<0.05$) NS priors.
$Q=M_{\rm BH}/M_{\rm NS}=1/q$, with its 90\% interval obtained from the reported mass ratio $q$.}
\renewcommand{\arraystretch}{1.5}   
\begin{ruledtabular}
\begin{tabular}{llcccc}
Event & NS prior & $m_{\rm BH}\,[M_\odot]$ & $m_{\rm NS}\,[M_\odot]$ & $\chi$ & $Q$ \\
\hline
\multirow{2}{*}{GW200105} & High spin & $8.9^{+1.2}_{-1.5}$ & $1.9^{+0.3}_{-0.2}$ & $0^{\,a}$ & $4.5^{+1.0}_{-1.2}$ \\
                          & Low spin  & $8.9^{+1.1}_{-1.3}$ & $1.9^{+0.2}_{-0.2}$ & $0^{\,a}$ & $4.8^{+1.1}_{-1.1}$ \\
\hline
\multirow{2}{*}{GW200115} & High spin & $5.7^{+1.8}_{-2.1}$ & $1.5^{+0.7}_{-0.3}$ & $-0.19^{+0.24}_{-0.50}$ & $3.8^{+2.4}_{-2.2}$ \\
                          & Low spin  & $5.9^{+1.4}_{-2.1}$ & $1.4^{+0.6}_{-0.2}$ & $-0.17^{+0.21}_{-0.56}$\,$^{b}$ & $4.2^{+2.1}_{-2.3}$ \\
\hline
GW230529                  & High spin & $3.6^{+0.8}_{-1.2}$ & $1.4^{+0.6}_{-0.2}$ & $-0.11^{+0.19}_{-0.35}$ & $2.6^{+1.1}_{-1.3}$ \\
\end{tabular}
\end{ruledtabular}
\begin{flushleft}
\footnotesize
$^a$\,$\chi_{1z}$ not reported; the BH spin is small and consistent with zero, so $\chi=0$ is adopted.\\
$^b$\,$\chi_{1z}$ not reported for the low-spin prior; estimated from $\chi_{\rm eff}$ as
$\chi\simeq\chi_{\rm eff}\,M/m_{\rm BH}$.
\end{flushleft}
\end{table*}
\begin{table*}[!t]
  \caption{$\Delta r=r_{\rm TD}-r_{\rm ISCO}$ for the BHNS events GW200105, GW200115 and GW230529,
    evaluated at the lower, middle (median) and upper values of the 90\% credible intervals of
    $M_{\rm BH}$, $M_{\rm NS}$ and $\chi$ for the high-spin and low-spin NS priors. For GW230529, the
    ``favoured'' case combines the lower bounds of the masses with the upper bound of the spin. For each
    case $R_{\rm NS}$ and $\eta$ are taken at that $M_{\rm NS}$ from the DDB and RMFNL EOSs.
    $\Delta r<0$ in every case, i.e., the sharp criterion $r_{\rm TD}\geq r_{\rm ISCO}$ is not satisfied
    for any configuration. Masses $m_{\rm BH}$ and $m_{\rm NS}$ are in $M_\odot$, and $r_{\rm ISCO}$,
    $r_{\rm TD}$, and $\Delta r$ in km.}
  \label{tab:dr_events}
  \begin{ruledtabular}
  \begin{tabular}{lllccccccccccc}
    & & & & & & & & \multicolumn{3}{c}{DDB} & \multicolumn{3}{c}{RMFNL} \\
    \cline{9-11}\cline{12-14}
    Event & NS prior & Case & $m_{\rm BH}$ & $m_{\rm NS}$ & $\chi$ & $Q$ & $r_{\rm ISCO}$ &
    $\eta$ & $r_{\rm TD}$ & $\Delta r$ & $\eta$ & $r_{\rm TD}$ & $\Delta r$ \\
    \hline
    \multirow{6}{*}{GW200105} & \multirow{3}{*}{High} & Lower  & 7.40 & 1.70 & $0.00$ & 4.35 & 65.56
      & 2.09 & 36.71 & $-28.85$ & 2.03 & 36.56 & $-29.01$ \\
     & & Middle & 8.90  & 1.90 & $0.00$ & 4.68 & 78.85 & 2.12 & 42.42 & $-36.44$ & 2.06 & 42.30 & $-36.55$ \\
     & & Upper  & 10.10 & 2.20 & $0.00$ & 4.59 & 89.48 & 2.29 & 46.76 & $-42.72$ & 2.23 & 46.68 & $-42.80$ \\
     & \multirow{3}{*}{Low} & Lower & 7.60 & 1.70 & $0.00$ & 4.47 & 67.33 & 2.09 & 37.51 & $-29.82$
      & 2.03 & 37.36 & $-29.97$ \\
     & & Middle & 8.90  & 1.90 & $0.00$ & 4.68 & 78.85 & 2.12 & 42.42 & $-36.44$ & 2.06 & 42.30 & $-36.55$ \\
     & & Upper  & 10.00 & 2.10 & $0.00$ & 4.76 & 88.60 & 2.20 & 46.59 & $-42.01$ & 2.15 & 46.50 & $-42.10$ \\
    \hline
    \multirow{6}{*}{GW200115} & \multirow{3}{*}{High} & Lower  & 3.60 & 1.20 & $-0.69$ & 3.00 & 43.13
      & 2.19 & 26.22 & $-16.92$ & 2.08 & 26.04 & $-17.09$ \\
     & & Middle & 5.70 & 1.50 & $-0.19$ & 3.80 & 55.62 & 2.11 & 31.99 & $-23.62$ & 2.03 & 31.83 & $-23.78$ \\
     & & Upper  & 7.50 & 2.20 & $0.05$  & 3.41 & 64.63 & 2.29 & 35.24 & $-29.39$ & 2.23 & 35.14 & $-29.49$ \\
     & \multirow{3}{*}{Low} & Lower & 3.80 & 1.20 & $-0.73$ & 3.17 & 46.22 & 2.19 & 27.36 & $-18.86$
      & 2.08 & 27.19 & $-19.04$ \\
     & & Middle & 5.90 & 1.40 & $-0.17$ & 4.21 & 57.11 & 2.13 & 32.91 & $-24.20$ & 2.04 & 32.74 & $-24.36$ \\
     & & Upper  & 7.30 & 2.00 & $0.03$  & 3.65 & 63.44 & 2.15 & 35.21 & $-28.23$ & 2.10 & 35.07 & $-28.37$ \\
    \hline
    \multirow{4}{*}{GW230529} & \multirow{4}{*}{High} & Lower & 2.40 & 1.20 & $-0.46$ & 2.00 & 26.35
      & 2.19 & 19.93 & $-6.41$ & 2.08 & 19.75 & $-6.59$ \\
     & & Middle   & 3.60 & 1.40 & $-0.11$ & 2.57 & 33.78 & 2.13 & 23.28 & $-10.50$ & 2.04 & 23.10 & $-10.68$ \\
     & & Upper    & 4.40 & 2.00 & $0.08$  & 2.20 & 37.27 & 2.15 & 23.68 & $-13.59$ & 2.10 & 23.50 & $-13.76$ \\
     & & Favoured & 2.40 & 1.20 & $0.08$  & 2.00 & 20.33 & 2.19 & 18.64 & $-1.69$  & 2.08 & 18.45 & $-1.88$ \\
  \end{tabular}
  \end{ruledtabular}
\end{table*}
However, the compactness alone does not determine the outcome. The monotonic and
nonmonotonic EOSs yield stars with nearly identical compactness at
$M_{\rm NS}=1.4\,M_\odot$ ($C\simeq0.184$), yet their disruption thresholds differ. For $M_{\rm NS}=1.4\,M_\odot$, the nonmonotonic NS is disrupted for BH spins as
low as $\chi_{\rm BH}\simeq-0.08$, whereas the monotonic NS requires
$\chi_{\rm BH}\gtrsim-0.04$. This difference arises from
their internal structure, quantified by $\eta=\rho_{\rm c}/\bar\rho$; a
larger $\eta$ corresponds to a more centrally condensed star. At fixed
compactness, the mass-shedding condition gives, at leading order in $\zeta$,
$r_{\rm TD}\propto\eta^{-1/3}$ [Eq.~(\ref{eq:rTD_1PN})], so a more centrally
condensed star is disrupted closer to the BH. The monotonic EOS
yields the larger value, $\eta=2.08$, and hence a smaller $r_{\rm TD}$, whereas
the nonmonotonic EOS, with $\eta=1.93$, yields a larger $r_{\rm TD}$. The
minimum spin required for disruption is set by the threshold condition
$r_{\rm TD}=r_{\rm ISCO}$. For a given BH mass, the ISCO radius decreases
monotonically with increasing spin. A larger $r_{\rm TD}$ is therefore matched by a larger
$r_{\rm ISCO}$, i.e., by a lower BH spin. Consequently, the nonmonotonic NS
reaches the threshold at a lower spin than the monotonic NS and can be disrupted
even for mildly retrograde BH spins.Therefore, while the compactness
characterizes the global properties of the NS, the parameter $\eta$ encodes its internal
structure and provides additional information on whether the NS is tidally disrupted or
plunges into the BH, which depends on the EOS of dense matter.

We applied the same analysis to the confirmed BHNS events GW200105 and GW200115~\cite{GW200105_115}
and a possible BHNS event GW230529~\cite{GW230529}. Their source properties are listed in
Table~\ref{tab:gw_events}, where the mass ratio is defined as
$Q=M_{\rm BH}/M_{\rm NS}=1/q$, with $q\leq1$ the mass ratio reported by the LVK
Collaboration. For each event we take the median values and the bounds of the 90\% credible intervals
of $M_{\rm BH}$, $M_{\rm NS}$, and $\chi_{\rm BH}$ reported by the LVK
Collaboration~\cite{GW200105_115,GW230529}, separately for the high-spin
and low-spin NS priors where available (Table~\ref{tab:gw_events}). We evaluate
$\Delta r=r_{\rm TD}-r_{\rm ISCO}$ for the DDB and RMFNL EOSs in three cases: the median
values of all parameters (``middle''), and the lower and upper bounds of their 90\%
credible intervals taken simultaneously for all parameters (``lower'' and ``upper''),
with $R$ and $\eta$ evaluated at the corresponding $M_{\rm NS}$. These bounds are
marginal (one-dimensional) intervals, and combining them ignores the correlations
between the parameters (e.g., $M_{\rm BH}$ and $M_{\rm NS}$ are anti-correlated through
the well-measured chirp mass). The lower and upper cases are therefore bracketing
configurations that span a range wider than the joint posterior, rather than individual
posterior samples. Since the lower bound of
$\chi_{\rm BH}$ corresponds to the most retrograde spin, which hinders disruption, the lower case is
not the configuration most favourable to disruption. For GW230529 we therefore also consider a
``favoured'' case, which combines the lower bounds of the masses with the upper bound of the spin.

The results are listed in Table~\ref{tab:dr_events}. For GW200105 and GW200115,
$\Delta r$ is strongly negative in every case, ranging from $\simeq-17$ to
$\simeq-43\,\mathrm{km}$, so the tidal radius lies well inside the ISCO and these
systems are clearly in the plunge regime for both EOSs. GW230529 is the most favourable
event for disruption, owing to its low BH mass ($Q\simeq2$--$2.6$). Its $\Delta r$ is
also negative in all cases, but much closer to zero: the lower case gives
$\Delta r=-6.41\,\mathrm{km}$ for DDB and $-6.59\,\mathrm{km}$ for RMFNL, and the favoured case gives
$\Delta r=-1.69\,\mathrm{km}$ and $-1.88\,\mathrm{km}$, respectively. The disruption criterion
therefore places all configurations of GW230529 in the plunge regime, but the favoured case lies within
$\simeq2\,\mathrm{km}$ of the threshold. Since the
criterion $r_{\rm TD}\geq r_{\rm ISCO}$ treats the onset of disruption as a sharp
threshold, whereas in practice mass shedding can begin shortly before the NS reaches
the ISCO, a small negative $\Delta r$ does not by itself exclude partial disruption.
The remnant-mass model, which is calibrated on NR simulations and does not impose the sharp threshold,
predicts a small but nonzero remnant mass for the lower and favoured cases of GW230529 (Sec.~\ref{sec:param_space-study}).
\subsection{Remnant disk mass}
\label{sec:rem_mass_results}

\begin{figure}[!t]
  \centering
  \includegraphics[width=\columnwidth]{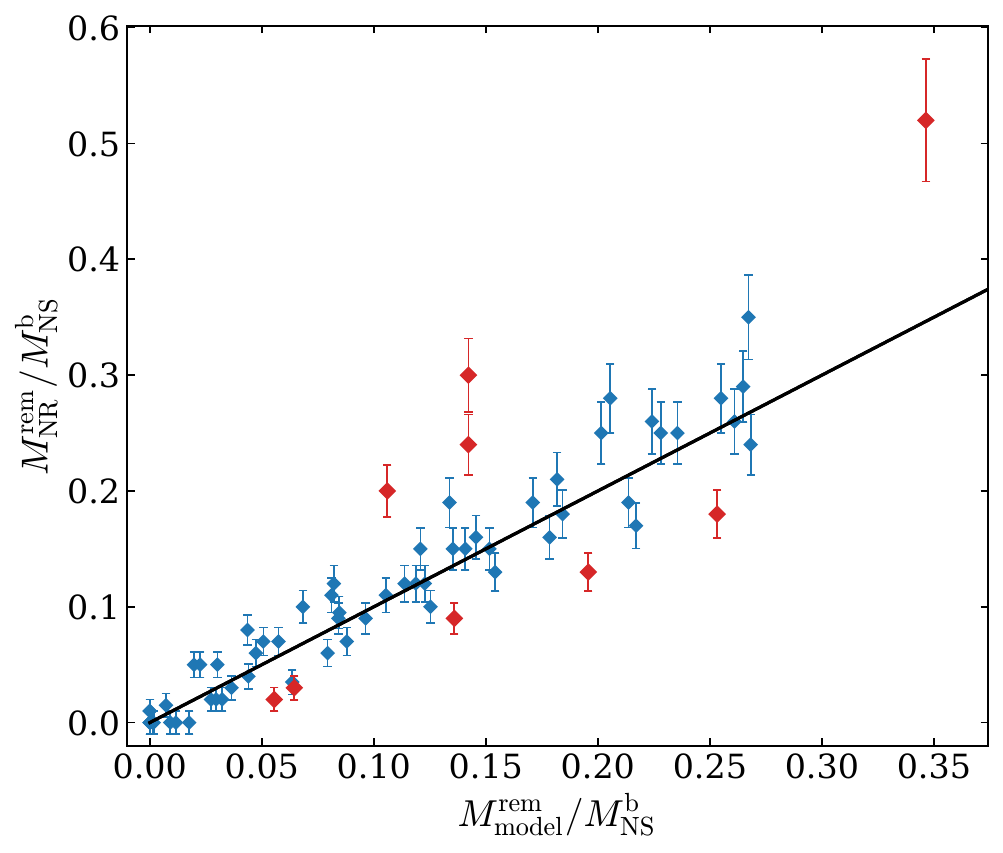}
  \caption{Comparison of the normalized remnant baryon mass predicted by the model,
    $\hat M^{\rm rem}_{\rm model}$, with the NR results $\hat M^{\rm rem}_{\rm NR}$, for the 70 BHNS
    simulations considered in this work. Error bars denote the NR uncertainties, and the black line marks
    perfect agreement, $\hat M^{\rm rem}_{\rm NR}=\hat M^{\rm rem}_{\rm model}$. Red points mark the nine
    simulations that deviate from the model by more than $3\,\Delta\hat M^{\rm rem}_{\rm NR}$.}
  \label{fig:Mrem_fit}
\end{figure}

We model the remnant disk mass remaining outside the BH after merger using
Eq.~\eqref{eq:Mrem_model}. The free parameters $(\alpha,\beta,\gamma,\delta)$ are determined through a
least-squares fit to the remnant masses $\hat M^{\rm rem}_{\rm NR}$ of 70 NR simulations, assigning to
each simulation the numerical error~\cite{Foucart2018}
\begin{equation}
  \Delta\hat M^{\rm rem}_{\rm NR}=\left[\left(0.1\,\hat M^{\rm rem}_{\rm NR}\right)^2+(0.01)^2\right]^{1/2},
  \label{eq:dM_NR}
\end{equation}
which combines a 10\% relative numerical error accounting for the limited ability of the simulations to resolve small remnant masses, and $1\%$ absolute error in the mass measurement. The best-fit parameters are $\alpha = 0.123 \pm 0.017$, $\beta = 0.162 \pm 0.008$, $\gamma = 0.635 \pm 0.064$, and $\delta = 1.830 \pm 0.254$.
Figure~\ref{fig:Mrem_fit} compares the model predictions with the NR results. The model reproduces the NR remnant masses over the full range $0\leq\hat M^{\rm rem}_{\rm NR}\lesssim0.5$. The goodness of fit is quantified by
\begin{equation}
  \chi^2 = \sum_{i=1}^{N} \left(\frac{\hat M^{\rm rem}_{{\rm NR},i} - \hat M^{\rm rem}_{{\rm model},i}}
  {\Delta \hat M^{\rm rem}_{{\rm NR},i}}\right)^2 ,
  \label{eq:chi2}
\end{equation}
together with the root-mean-square (rms) deviation
\begin{equation}
  \sigma_{\rm rms} = \left[\frac{1}{N}\sum_{i=1}^{N}
  \left(\hat M^{\rm rem}_{{\rm NR},i} - \hat M^{\rm rem}_{{\rm model},i}\right)^2\right]^{1/2}.
  \label{eq:rms}
\end{equation}
The two quantities measure different aspects of the fit. $\chi^2$ weights each deviation by the
uncertainty of the corresponding simulation and thus tests whether the model is consistent with the
NR error bars; it is dominated by configurations with small remnant masses, for which these error
bars are small. $\sigma_{\rm rms}$ instead treats all simulations equally and gives the typical size
of the error in the predicted remnant mass, in units of the NS baryon mass $M_{\rm NS}^{\rm b}$.

As a reference, we fit the FF18 model~\cite{Foucart2018},
$\hat M^{\rm rem}=[\max(\alpha\,(1-2C)/\nu^{1/3}-\beta\,\hat r_{\rm ISCO}\,C/\nu+\gamma,\,0)]^{\delta}$,
to the same simulations, minimizing $\chi^2$ with the same uncertainties; the resulting coefficients
are consistent with the published ones. For all $N=70$ simulations ($N-4=66$ degrees of freedom),
our model gives $\chi^2=246$, i.e., $\chi^2_{\rm red}\equiv\chi^2/(N-4)=3.7$ and
$\sigma_{\rm rms}=0.043$, while FF18 gives $\chi^2=137$, $\chi^2_{\rm red}=2.1$, and
$\sigma_{\rm rms}=0.029$. In terms of the remnant mass itself, $\sigma_{\rm rms}=0.043$
means that for a typical simulation our model predicts a remnant mass that differs from the NR value
by about $4\%$ of the NS baryon mass, i.e., by $\simeq0.07\,M_\odot$ for
$M_{\rm NS}^{\rm b}\simeq1.55\,M_\odot$, whereas the typical difference for FF18 is about $3\%$
($\simeq0.04\,M_\odot$).
The difference between the two models is concentrated in a few configurations: nine simulations
deviate from our model by more than $3\,\Delta\hat M^{\rm rem}_{\rm NR}$ (red points in
Fig.~\ref{fig:Mrem_fit}) and alone contribute $53\%$ of its $\chi^2$. Removing them and refitting both
models to the remaining 61 simulations ($57$ degrees of freedom) gives $\chi^2_{\rm red}=1.96$
($\sigma_{\rm rms}=0.024$) for our model and $1.52$ ($\sigma_{\rm rms}=0.020$) for FF18. The typical errors of the two models then
differ by only $\simeq0.4\%$ of $M_{\rm NS}^{\rm b}$ ($\simeq0.006\,M_\odot$), well below the NR
uncertainty, so outside these nine configurations the two models describe the NR data with
essentially the same accuracy. The remaining $\chi^2_{\rm red}>1$ of both suggests that
Eq.~\eqref{eq:dM_NR} underestimates the true NR uncertainty, e.g., by not including systematic
differences between evolution codes.

Although the purely phenomenological FF18 fit reproduces the NR data with a somewhat lower $\chi^2$,
the present model attains an accuracy comparable to the NR uncertainties with a functional form
derived from the physical condition for tidal disruption rather than chosen empirically. This
provides the following advantages. (i) Its first
term is the tidal-disruption radius $r_{\rm TD}/R_{\rm NS}$, obtained by balancing the tidal field of
the Kerr BH against the self-gravity of the NS (ii) The BH spin enters the tidal term
itself, through the tidal field of the Kerr metric, whereas in FF18 the spin appears only through the
ISCO radius and the tidal term depends on the binary only through $\nu$. (iii) The EOS enters through
the internal structure of the star, via $\eta$, and not through the compactness alone, so that NSs
with the same radius but different density profiles yield different remnant masses. Our model with the same number of free parameters, predicts the NR remnant masses to within about $4\%$ of the
NS baryon mass shows that the tidal-disruption radius captures the main physics of remnant formation. These nine NR results fall into two physically distinct groups:
\begin{itemize}
  \item \textit{Underestimated remnant mass.} Four simulations of $\Gamma=2$ polytropes with
  rapidly spinning BHs ($\chi_{\rm BH}\geq0.9$) lie above the line
  $\hat M^{\rm rem}_{\rm NR}=\hat M^{\rm rem}_{\rm model}$, with deviations of up to
  $0.17\,M_{\rm NS}^{\rm b}$. These stars are strongly centrally condensed
  ($\eta\simeq3.8$--$3.9$, compared with $\eta\simeq2.0$--$2.5$ for the nuclear-physics-motivated
  EOSs), so that the mass-shedding condition places $r_{\rm TD}$ close to the BH. In a
  $\Gamma=2$ polytrope, however, the large $\eta$ is associated with an extended, low-density
  envelope that is weakly bound and is stripped more easily than the single parameter $\eta$
  implies. The model therefore underestimates the matter remaining outside the BH, most strongly at
  high spin, where the disruption occurs far outside the ISCO and the remnant mass is most sensitive
  to the outer layers of the star.

    \item \textit{Overestimated remnant mass.} Five simulations with piecewise-polytropic EOSs at
  $\chi_{\rm BH}=0.75$ and $Q=3$--$5$ lie below the line, with deviations of
  $0.03$--$0.07\,M_{\rm NS}^{\rm b}$. Three of them are the most compact configurations of this set
  ($C\simeq0.18$), which lie close to the disruption threshold and have small NR remnant
  masses ($\hat M^{\rm rem}_{\rm NR}\leq0.09$). Near the threshold the remnant mass decreases steeply
  with compactness, so that a small offset in the predicted onset of disruption produces a large
  relative error. The remaining
  two ($Q=3$) have larger remnant masses, $0.13$ and $0.18\,M_{\rm NS}^{\rm b}$, and are
  overestimated by $\simeq0.07\,M_{\rm NS}^{\rm b}$. For these EOSs, $\eta$ varies only weakly
  ($\eta\simeq2.0$--$2.2$), so the tidal term distinguishes them mainly through the compactness, and
  differences in their internal structure beyond $\eta$ are not captured by the model.
\end{itemize}

\begin{figure}[!t]
  \centering
  \includegraphics[width=\columnwidth]{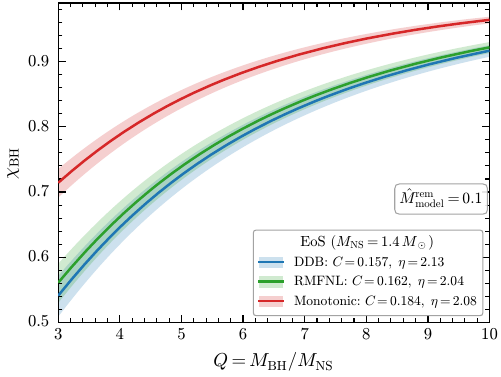}
  \caption{Contours of constant remnant mass, $\hat M^{\rm rem}_{\rm model}=0.1\,M_{\rm NS}^{\rm b}$,
    predicted by Eq.~\eqref{eq:Mrem_model}. The shaded bands mark
    $0.09\leq\hat M^{\rm rem}_{\rm model}/M_{\rm NS}^{\rm b}\leq0.11$ and indicate the sensitivity of the
    contours to a change of $\pm0.01\,M_{\rm NS}^{\rm b}$ in the remnant mass. Binaries above (to the
    upper left of) each curve leave $\hat M^{\rm rem}_{\rm model}>0.1\,M_{\rm NS}^{\rm b}$ outside the
    BH. More compact NSs require higher prograde BH spins to retain the same remnant mass, and the
    required spin increases with $Q$; at $Q=3$, for instance, it ranges from $\chi_{\rm BH}\simeq0.54$
    for DDB to $\simeq0.72$ for the monotonic EOS.}
  \label{fig:Mrem01_contours}
\end{figure}

For the nuclear-physics EOSs the model shows no significant systematic bias, with a mean offset
$\langle\hat M^{\rm rem}_{\rm NR}-\hat M^{\rm rem}_{\rm model}\rangle=0.001\,M_{\rm NS}^{\rm b}$, whereas
for the $\Gamma=2$ polytropes it underestimates the remnant mass by $0.064\,M_{\rm NS}^{\rm b}$ on
average.
Figure~\ref{fig:Mrem01_contours} shows the contours of
$\hat M^{\rm rem}_{\rm model}=0.1\,M_{\rm NS}^{\rm b}$ predicted by our model in the
$(Q,\chi_{\rm BH})$ plane for $M_{\rm NS}=1.4\,M_\odot$ stars described by DDB, RMFNL, and Monotonic
EOSs with different compactness, $C=0.157$, $0.162$, and $0.184$ ($R_{\rm NS}=13.14$, $12.78$, and
$11.25\,\mathrm{km}$). The formation of massive disks is favored by low mass ratios and high prograde BH spins. The contours for the DDB and RMFNL EOSs nearly coincide, with their shaded bands overlapping over
the entire range of $Q$. In our model, the remnant mass is set by the competition between two
quantities, both measured in units of the NS radius: the tidal-disruption radius
$r_{\rm TD}/R_{\rm NS}$, which depends on $C$ and $\eta$, and the ISCO term
$\hat r_{\rm ISCO}\,\rho/\nu$, which depends on $\rho$. A larger $r_{\rm TD}/R_{\rm NS}$ increases the
remnant mass, while a larger ISCO term decreases it. Compared with the DDB star, the RMFNL star has a
higher compactness ($C=0.162$ versus $0.157$) and a lower central condensation ($\eta=2.04$ versus
$2.13$), which together increase $r_{\rm TD}/R_{\rm NS}$ by $\simeq2\%$. At the same time, its
larger $\rho$ ($0.162$ versus $0.158$) increases the ISCO term by $\simeq2.5\%$. Since both terms
change by nearly the same amount, their difference, and hence the remnant mass, remains almost
unchanged. The RMFNL star therefore requires only a marginally higher BH spin, by
$\Delta\chi_{\rm BH}\simeq0.02$ at $Q=3$ and $\lesssim0.01$ for $Q\gtrsim7$. The remnant mass model combines a physically motivated description of tidal disruption,
based on the general-relativistic mass-shedding condition and the Kerr ISCO, with
coefficients calibrated against NR simulations. The calibration set spans a wide range of
mass ratios, BH spins, and NS compactnesses, so that the model is expected to provide
reliable estimates of the remnant mass over most of the astrophysically relevant parameter
space.

The accuracy of the model is highest for NSs described by nuclear-physics-motivated EOSs,
for which it reproduces the NR results without significant systematic bias and with a
typical scatter of $\simeq0.03\,M_{\rm NS}^{\rm b}$. We note that $\Gamma=2$ polytropes are idealized EOSs, commonly adopted in NR simulations
for their simplicity, but they lack a crust and the density-dependent stiffness of
nuclear-physics models~\cite{Read2009,LattimerPrakash2016}. In particular, they are not
directly supported by observations of massive pulsars~\cite{Antoniadis2013,Fonseca2021} or by
the radius and tidal deformability constraints from NICER~\cite{Miller_2019} and GW170817~\cite{GW170817}, which are naturally
accommodated by nuclear-physics-motivated EOSs. The reduced accuracy of the model for
polytropic stars is therefore of limited relevance for astrophysical BHNS systems.
\subsection{Parameter space study}
\label{sec:param_space-study}
\begin{figure*}[ht]
  \centering
  \includegraphics[width=0.32\textwidth]{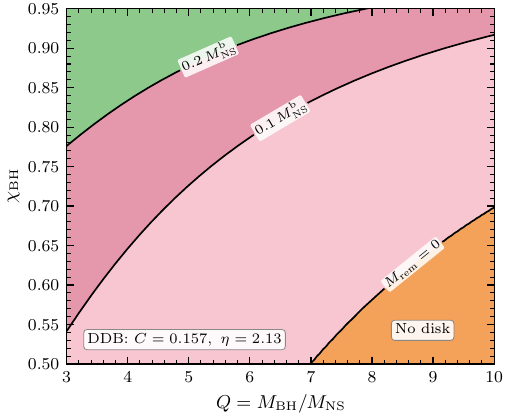}\hfill
  \includegraphics[width=0.32\textwidth]{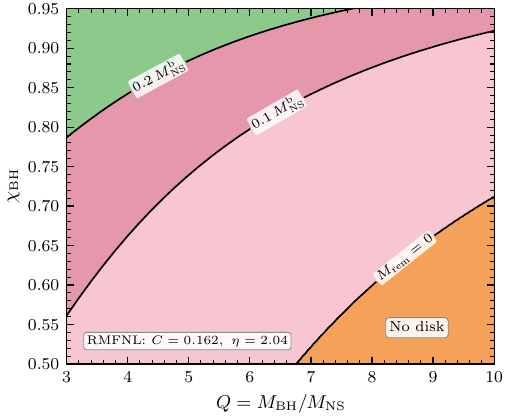}\hfill
  \includegraphics[width=0.32\textwidth]{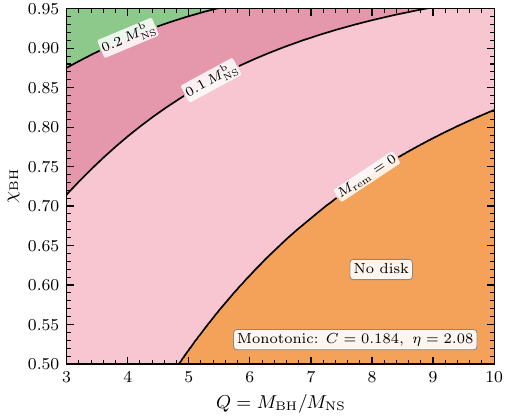}
  \caption{Contours of the model remnant mass, $M_{\rm rem}^{\rm model}=0$, $0.1$, and
    $0.2\,M_{\rm NS}^{\rm b}$, for
    $1.4\,M_\odot$ NSs described by the DDB (left; $C=0.157$, $\eta=2.13$,
    $R_{\rm NS}=13.14\,\mathrm{km}$), RMFNL (middle; $C=0.162$, $\eta=2.04$,
    $R_{\rm NS}=12.78\,\mathrm{km}$), and monotonic (right; $C=0.184$, $\eta=2.08$,
    $R_{\rm NS}=11.25\,\mathrm{km}$) EOSs. The shaded regions mark the parameter space in
    which no matter remains outside the BH (orange),
    $0<M_{\rm rem}^{\rm model}<0.1\,M_{\rm NS}^{\rm b}$ (pink),
    $0.1<M_{\rm rem}^{\rm model}<0.2\,M_{\rm NS}^{\rm b}$ (rose), and
    $M_{\rm rem}^{\rm model}>0.2\,M_{\rm NS}^{\rm b}$ (green). The DDB and RMFNL stars yield
    nearly identical regions, whereas the more compact monotonic star requires
    substantially higher BH spins to form a remnant disk.}
  \label{fig:disk_contours}
\end{figure*}
\begin{figure}[t]
  \centering
  \includegraphics[width=\columnwidth]{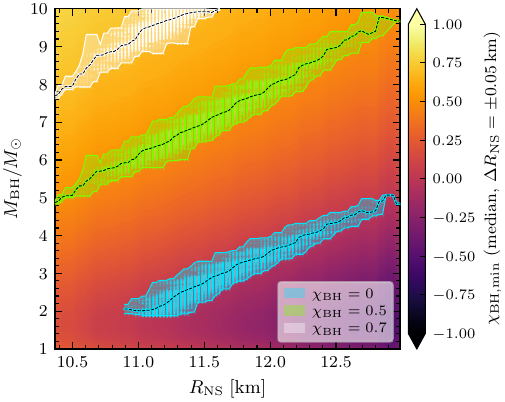}
  \caption{Tidal disruption threshold, $M_{\rm rem}^{\rm model} = 0$, of a
    $M_{\rm NS} = 1.4\,M_\odot$ NS in the $R_{\rm NS}$--$M_{\rm BH}$ plane, predicted by
    Eq.~\eqref{eq:Mrem_model} for the $10\,077$ EOSs of~\cite{asim_kumar_saha_2026,Verma:2025dez}. For aligned BH
    spins $\chi_{\rm BH} = 0$ (cyan), $0.5$ (green), and $0.7$ (white), the faint dots show
    the threshold BH mass of the individual EOSs, the shaded bands its range over the EOSs
    within $\Delta R_{\rm NS} = \pm 0.05\,\mathrm{km}$ of each sampled radius, and the dashed
    lines its median. The color map shows the median, over the same EOSs, of the minimum
    aligned BH spin $\chi_{\rm BH,min}$ required for disruption.}
  \label{fig:chi_min_EoS}
\end{figure}
\begin{figure}[!t]
  \centering
  \vspace{4pt}
  \includegraphics[width=\columnwidth]{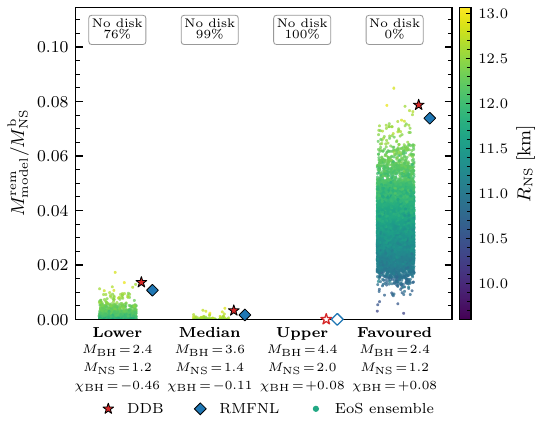}
  \caption{Remnant mass $\hat M^{\rm rem}_{\rm model}/M_{\rm NS}^{\rm b}$ predicted by
    Eq.~\eqref{eq:Mrem_model} for GW230529 in three configurations: the median values of $M_{\rm BH}$,
    $M_{\rm NS}$, and $\chi_{\rm BH}$, and the lower and upper bounds of their marginal 90\% credible
    intervals taken simultaneously for all parameters (values listed below each column; see
    Table~\ref{tab:gw_events}). Dots show the individual EOSs of the $10\,077$-member ensemble that yield
    a remnant disk ($\hat M^{\rm rem}_{\rm model}>0$), colored by the radius $R_{\rm NS}$ of the
    corresponding NS. The percentage of EOSs that yield no disk ($\hat M^{\rm rem}_{\rm model}=0$) is
    given at the top of each column. The star and diamond denote the DDB and RMFNL EOSs, respectively;
    open symbols at $\hat M^{\rm rem}_{\rm model}=0$ indicate that no disk is formed.}
  \label{fig:GW230529_Mrem}
\end{figure}
The remnant mass model presented in Sec.~\ref{sec:Mrem_model} provides a simple way to estimate the region of parameter space in which the NS is tidally disrupted, and to estimate how much matter remains outside the BH at late times. Such predictions are essential for identifying which BHNS mergers can power short gamma-ray bursts (SGRBs), since the formation of a sufficiently massive accretion disk around the remnant BH is commonly regarded as a necessary condition for launching a relativistic jet. Figure~\ref{fig:disk_contours} shows the predicted remnant mass for $M_{\rm NS} = 1.4\,M_\odot$ stars described by DDB, RMFNL, and Monotonic EOSs with compactness ($C = 0.157$, $0.162$, and $0.184$). 

\begin{table*}[ht]
\caption{Remnant mass $\hat M^{\rm rem}_{\rm model}$, in units of the NS baryon mass
    $M_{\rm NS}^{\rm b}$, predicted by Eq.~\eqref{eq:Mrem_model} for GW230529, evaluated at the lower
    bound, median, and upper bound of the 90\% credible intervals of the binary parameters:
    $(M_{\rm BH},M_{\rm NS})=(2.4,1.2)$, $(3.6,1.4)$, and $(4.4,2.0)\,M_\odot$, with the aligned BH spin
    $\chi_{\rm BH}$ as listed. The favoured case uses the lower-bound masses, $(2.4,1.2)\,M_\odot$, with
    the upper-bound spin. For the ensemble of $10\,077$ EOSs, $N_{\rm disk}/N$ is the number of EOSs with
    $\hat M^{\rm rem}_{\rm model}>0$ out of the $N$ EOSs that support a NS of mass $M_{\rm NS}$; a value
    of 0 denotes no remnant disk.}
  \label{tab:Mrem_GW230529}
  \begin{ruledtabular}
  \begin{tabular}{lcccccccc}
 & & \multicolumn{2}{c}{$M_{\rm rem}^{\rm model}$} & \multicolumn{5}{c}{$10\,077$ EoSs} \\
\cline{3-4} \cline{5-9}
 & & & & & \multicolumn{4}{c}{EoS with largest $M_{\rm rem}^{\rm model}$} \\
\cline{6-9}
Bound & $\chi_{\rm BH}$ & DDB & RMFNL & $N_{\rm disk}/N$
      & $R_{\rm NS}\,[\mathrm{km}]$ & $\eta$ & $\rho$ & $M_{\rm rem}^{\rm model}$ \\
\hline
Lower   & $-0.46$ & 0.014 & 0.011 & 2383/10077  & 12.836 & 2.236 & 0.131 & 0.017 \\
Median  & $-0.11$ & 0.003 & 0.002 & 100/10077   & 12.884 & 2.258 & 0.157 & 0.004 \\
Upper   & $+0.08$ & 0     & 0     & 0/8921      & 13.028 & 1.706 & 0.245 & 0     \\
Favored & $+0.08$ & 0.079 & 0.074 & 10077/10077 & 12.836 & 2.236 & 0.131 & 0.085 \\
\end{tabular}
\end{ruledtabular}
\end{table*}

The remnant mass depends strongly on the NS compactness and, to a lesser extent, on
$\eta$. For the DDB and RMFNL stars, the region in which no disk forms is confined to large
mass ratios and moderate BH spins. For the
more compact monotonic star ($C=0.184$), it extends to progressively lower mass ratios and
higher BH spins. This behavior reflects the competition
between the tidal disruption radius and the ISCO: a more compact, more strongly
self-bound NS is disrupted closer to the BH, so that disruption outside the ISCO, and hence
the formation of a remnant disk, requires a lower mass ratio or a higher prograde BH spin,
which reduces the ISCO radius. Correspondingly, the regions with
$M_{\rm rem}^{\rm model}>0.1\,M_{\rm NS}^{\rm b}$ and
$M_{\rm rem}^{\rm model}>0.2\,M_{\rm NS}^{\rm b}$ shrink toward low $Q$ and high
$\chi_{\rm BH}$, and the largest remnant mass in the plane decreases from
$\simeq0.42\,M_{\rm NS}^{\rm b}$ for DDB and RMFNL to $\simeq0.37\,M_{\rm NS}^{\rm b}$ for the
monotonic star. The DDB and RMFNL stars yield nearly identical regions: although RMFNL is
slightly more compact ($C=0.162$ versus $0.157$), it is also less centrally condensed
($\eta=2.04$ versus $2.13$), and the two effects largely compensate each other. Since both
$C$ and $\eta$ are determined by the EOS of dense matter, these results imply that the
fraction of BHNS mergers capable of producing SGRBs, and thus the expected SGRB rate from
BHNS systems, is highly sensitive to the nuclear EOS.


Figure~\ref{fig:chi_min_EoS} shows the disruption threshold predicted by the remnant mass
model of Eq.~\eqref{eq:Mrem_model} for a set of $10\,077$ phenomenological
(model-agnostic) EOSs~\cite{asim_kumar_saha_2026,Verma:2025dez}. For each EOS we consider a NS of fixed gravitational
mass $M_{\rm NS} = 1.4\,M_\odot$, with its radius $R_{\rm NS}$, density ratio $\eta$, and
determine the threshold for tidal disruption by solving $\hat M_{\rm rem}^{\rm model} = 0$ using Eq.~\ref{eq:Mrem_model}.
The range of radii spanned by the ensemble,
$10.37\,\mathrm{km} \le R_{\rm NS} \le 12.99\,\mathrm{km}$, is sampled at $100$ equally
spaced values, and at each sampled radius we select all EOSs whose $1.4\,M_\odot$ star has
a radius within $\pm 0.05\,\mathrm{km}$ of that value.

For a fixed BH spin, every EOS yields a single threshold BH mass, $M_{\rm BH}^{\rm th}$,
above which the NS is swallowed by the BH without leaving matter outside it
($\hat M^{rem}_{ model} = 0$), and below which it is tidally disrupted ($\hat M^{rem}_{model} > 0$).  The vertical strips in
Fig.~\ref{fig:chi_min_EoS} show the spread of threshold black hole mass $M_{\rm BH}^{\rm th}$ 
among the EOSs in each radius bin, and the dashed lines its median. Since these EOSs predict essentially the same
radius but differ in their internal structure, as encoded in $\eta$, the
width of each strip quantifies the dependence of the disruption threshold on the NS
properties: below a strip the NS is disrupted for all EOSs in the bin,
above it for none, and within it the outcome depends on the EOS.
The background color map shows the median, over the EOSs in each radius bin, of the
minimum aligned BH spin $\chi$ required for
disruption: a binary leaves a remnant disk if $\chi_{\rm BH} \ge \chi_{\rm BH,min}$, so
that each band traces the contour $\chi_{\rm BH,min} = \chi_{\rm BH}$. Negative values of
$\chi_{\rm BH,min}$, found below the $\chi_{\rm BH} = 0$ band, indicate that the NS is
disrupted even by a moderately retrograde BH, whereas above it a prograde spin is
required. For instance, for $R_{\rm NS} \simeq 11.5\,\mathrm{km}$ a nonspinning BH
disrupts the NS only for $M_{\rm BH} \lesssim 2.9\,M_\odot$, while a $4\,M_\odot$ BH
requires $\chi_{\rm BH} \gtrsim 0.16$. Increasing the prograde BH spin shifts the threshold
to larger BH masses, because it moves the ISCO inward: the threshold rises from
$M_{\rm BH} \approx 2$--$5\,M_\odot$ for $\chi_{\rm BH} = 0$ to
$M_{\rm BH} \approx 5$--$10\,M_\odot$ for $\chi_{\rm BH} = 0.5$ across the radius range. For a given BH
spin, NSs with larger radii, i.e., stiffer EOSs, can be disrupted by more massive BHs,
while for $R_{\rm NS} \lesssim 10.9\,\mathrm{km}$ a nonspinning BH does not disrupt the NS
for any BH mass considered here. 
Using the disruption threshold, the EOS ensemble can be used to identify
the subset of EOSs compatible with a given binary configuration. For a fixed point in the
$R_{\rm NS}$--$M_{\rm BH}$ plane, the EOSs whose $1.4\,M_\odot$ star lies within
$\Delta R_{\rm NS} = \pm 0.05\,\mathrm{km}$ of the chosen radius yield a distribution of
minimum BH spins, $\chi_{\rm BH,min}$, required for tidal disruption. Conversely, an
independent constraint on the BH spin, or on the presence or absence of an
electromagnetic counterpart, selects the subset of EOSs consistent with the observed
outcome. 

In Sec.~\ref{sec:TD_or_plunge} we assessed whether the NSs in GW200105, GW200115, and GW230529
were tidally disrupted or plunged directly into the BH, using the difference between the
tidal disruption radius and the ISCO radius, $\Delta r = r_{\rm TD} - r_{\rm ISCO}$
(Table~\ref{tab:dr_events}). 
For GW200105 and GW200115, $\Delta r$ is strongly negative
across the entire credible range of the binary parameters, indicating that the NS
crosses the ISCO before it can be disrupted; these systems are therefore expected to
plunge without leaving a remnant disk. GW230529 is the only event for which $\Delta r$
approaches zero within its credible range, making it the most promising candidate for
partial tidal disruption. 
We therefore apply the remnant mass model of
Eq.~\eqref{eq:Mrem_model} to GW230529, describing the NS with the DDB and RMFNL EOSs as
well as with the ensemble of $10\,077$ phenomenological EOSs. 
As shown in Fig.~\ref{fig:GW230529_Mrem} and summarized in Table~\ref{tab:Mrem_GW230529}, the
model predicts at most a small remnant disk for GW230529, and only toward the low-mass end of
the credible interval. At the lower bound ($M_{\rm BH}=2.4\,M_\odot$, $M_{\rm NS}=1.2\,M_\odot$,
$\chi_{\rm BH}=-0.46$), both DDB and RMFNL yield $M_{\rm rem}^{\rm model}\simeq0.01\,M_{\rm NS}^{\rm b}$,
and about $24\%$ of the EOS ensemble ($2383$ of $10\,077$) predicts disk formation, with
$M_{\rm model}^{\rm rem}\le0.017\,M_{\rm NS}^{\rm b}$. At the median values, the remnant mass
decreases to $\simeq0.002$--$0.003\,M_{\rm NS}^{\rm b}$ for DDB and RMFNL, and only $1\%$ of the
ensemble ($100$ EOSs) forms a disk, with $M_{\rm model}^{\rm rem}\le0.004\,M_{\rm NS}^{\rm b}$.
At the upper bound, none of the $8921$ EOSs supporting a $2\,M_\odot$ NS, including DDB and
RMFNL, leads to a remnant disk, and even the EOS closest to the threshold lies well within
the plunge regime. The EOSs that do form a disk are those yielding the largest, i.e., least
compact, NSs at a given mass ($R_{\rm NS}\gtrsim12.8\,\mathrm{km}$), as indicated by the color
coding in Fig.~\ref{fig:GW230529_Mrem}. Overall, these results suggest that GW230529 most likely ended in a plunge: for the median parameters the
remnant mass is $\lesssim0.004\,M_{\rm NS}^{\rm b}$ for all EOSs. A remnant of up to
$\simeq0.08\,M_{\rm NS}^{\rm b}$ is possible only if both component masses lie near the lower end of
their credible intervals and the BH spin is prograde. Since these bounds are taken from marginal
intervals, a quantitative statement requires evaluating the model over the joint posterior samples of
the event.
\section{Summary and conclusions}
\label{sec:conclusion}

Using the stretching eigenvalue of the relativistic tidal tensor in the Kerr
spacetime~\cite{Marck1983,XinMummery2026}, evaluated on circular equatorial orbits, we
formulated a relativistic mass-shedding condition for a NS orbiting a Kerr BH by balancing
the tidal acceleration at the stellar surface against the self-gravity of the star. The
resulting condition, Eq.~(\ref{eq:TD_threshold_NS}), reduces to the Newtonian tidal radius in
the weak-field limit, but for BHNS binaries the gravitational field at the onset of
disruption is strong, and the relativistic tidal radius exceeds the Newtonian estimate
significantly; for $Q=3$, $C=0.15$, $\eta=2.1$, and $\chi=0.8$,
for which disruption occurs outside the ISCO, the difference is about 11\%.

Using the relativistic tidal radius in place of the phenomenological disruption term of the FF18 fit,
we constructed a remnant-mass model with the same number of free parameters. In contrast to the fitting
formulas of FF12~\cite{Foucart2012} and FF18, the model accounts for the BH spin in the tidal term and
for the density profile of the NS through $\eta$. Calibrated on 70 NR simulations, it reproduces the
remnant mass with an accuracy comparable to FF18, with a typical scatter of
$\simeq0.03\,M_{\rm NS}^{\rm b}$ for nuclear-physics-motivated EOSs; the largest deviations occur for
idealized $\Gamma=2$ polytropes. Applying the model to the DDB, RMFNL, monotonic, and nonmonotonic EOSs
and to an ensemble of $10\,077$ phenomenological EOSs, we found that the disruption threshold depends
primarily on the NS compactness, with $\eta$ introducing an additional dependence at fixed
compactness. Since only BHNS mergers leaving a sufficiently massive accretion disk are expected to power
SGRBs, this implies that the fraction of BHNS mergers producing SGRBs is sensitive to the nuclear EOS.
Among the observed events, GW200105 and GW200115 lie well within the plunge regime, whereas GW230529
lies close to the disruption threshold and most likely also plunged: for its median parameters the
predicted remnant mass is $\hat M^{\rm rem}_{\rm model}\lesssim0.004$ for all EOSs considered, and a
remnant of up to $\simeq0.08$ is possible only if both masses lie near the lower end of their credible
intervals and the BH spin is prograde.

Future work should extend the analysis to misaligned BH spins and spinning NSs, evaluate the model over
the full posterior distributions of observed events, and include new NR simulations at smaller mass
ratios and for a broader range of EOSs to refine the calibration. Combined with population models and
electromagnetic observations of BHNS mergers, the model can then be used to estimate the fraction of
systems powering SGRBs and kilonovae and to constrain the EOS of dense matter.
\section*{Acknowledgements}
\label{sec:acknowledgements}
The authors thank the Indian Institute of Science Education and Research (IISER) Bhopal for
providing all the research and infrastructure facilities. Y.C. acknowledges the financial assistance
received through the Junior Research Fellowship (JRF), awarded by the Council of Scientific and
Industrial Research (CSIR), Government of India (File No:	09/1020(24279)/2025-EMR-I). The authors also extend their appreciation to A.K. Saha for insightful discussions, which contributed to this work.

\bibliography{References}

\begin{thebibliography}{71}%
\makeatletter
\providecommand \@ifxundefined [1]{%
 \@ifx{#1\undefined}
}%
\providecommand \@ifnum [1]{%
 \ifnum #1\expandafter \@firstoftwo
 \else \expandafter \@secondoftwo
 \fi
}%
\providecommand \@ifx [1]{%
 \ifx #1\expandafter \@firstoftwo
 \else \expandafter \@secondoftwo
 \fi
}%
\providecommand \natexlab [1]{#1}%
\providecommand \enquote  [1]{``#1''}%
\providecommand \bibnamefont  [1]{#1}%
\providecommand \bibfnamefont [1]{#1}%
\providecommand \citenamefont [1]{#1}%
\providecommand \href@noop [0]{\@secondoftwo}%
\providecommand \href [0]{\begingroup \@sanitize@url \@href}%
\providecommand \@href[1]{\@@startlink{#1}\@@href}%
\providecommand \@@href[1]{\endgroup#1\@@endlink}%
\providecommand \@sanitize@url [0]{\catcode `\\12\catcode `\$12\catcode `\&12\catcode `\#12\catcode `\^12\catcode `\_12\catcode `\%12\relax}%
\providecommand \@@startlink[1]{}%
\providecommand \@@endlink[0]{}%
\providecommand \url  [0]{\begingroup\@sanitize@url \@url }%
\providecommand \@url [1]{\endgroup\@href {#1}{\urlprefix }}%
\providecommand \urlprefix  [0]{URL }%
\providecommand \Eprint [0]{\href }%
\providecommand \doibase [0]{https://doi.org/}%
\providecommand \selectlanguage [0]{\@gobble}%
\providecommand \bibinfo  [0]{\@secondoftwo}%
\providecommand \bibfield  [0]{\@secondoftwo}%
\providecommand \translation [1]{[#1]}%
\providecommand \BibitemOpen [0]{}%
\providecommand \bibitemStop [0]{}%
\providecommand \bibitemNoStop [0]{.\EOS\space}%
\providecommand \EOS [0]{\spacefactor3000\relax}%
\providecommand \BibitemShut  [1]{\csname bibitem#1\endcsname}%
\let\auto@bib@innerbib\@empty
\bibitem [{\citenamefont {Abbott}\ \emph {et~al.}(2021)\citenamefont {Abbott} \emph {et~al.}}]{GW200105_115}%
  \BibitemOpen
  \bibfield  {author} {\bibinfo {author} {\bibfnamefont {R.}~\bibnamefont {Abbott}} \emph {et~al.} (\bibinfo {collaboration} {LIGO Scientific, KAGRA, and Virgo}),\ }\bibfield  {title} {\bibinfo {title} {Observation of gravitational waves from two neutron star--black hole coalescences},\ }\href {https://doi.org/10.3847/2041-8213/ac082e} {\bibfield  {journal} {\bibinfo  {journal} {Astrophys. J. Lett.}\ }\textbf {\bibinfo {volume} {915}},\ \bibinfo {pages} {L5} (\bibinfo {year} {2021})},\ \Eprint {https://arxiv.org/abs/2106.15163} {arXiv:2106.15163 [astro-ph.HE]} \BibitemShut {NoStop}%
\bibitem [{\citenamefont {Abac}\ \emph {et~al.}(2024)\citenamefont {Abac} \emph {et~al.}}]{GW230529}%
  \BibitemOpen
  \bibfield  {author} {\bibinfo {author} {\bibfnamefont {A.~G.}\ \bibnamefont {Abac}} \emph {et~al.} (\bibinfo {collaboration} {LIGO Scientific, Virgo, and KAGRA}),\ }\bibfield  {title} {\bibinfo {title} {Observation of gravitational waves from the coalescence of a $2.5$--$4.5\,m_\odot$ compact object and a neutron star},\ }\href {https://doi.org/10.3847/2041-8213/ad5beb} {\bibfield  {journal} {\bibinfo  {journal} {Astrophys. J. Lett.}\ }\textbf {\bibinfo {volume} {970}},\ \bibinfo {pages} {L34} (\bibinfo {year} {2024})},\ \Eprint {https://arxiv.org/abs/2404.04248} {arXiv:2404.04248 [astro-ph.HE]} \BibitemShut {NoStop}%
\bibitem [{\citenamefont {{LIGO Laboratory}}(2026)}]{LIGOwebsite}%
  \BibitemOpen
  \bibfield  {author} {\bibinfo {author} {\bibnamefont {{LIGO Laboratory}}},\ }\href@noop {} {\bibinfo {title} {{LIGO} -- laser interferometer gravitational-wave observatory}},\ \bibinfo {howpublished} {\url{https://www.ligo.caltech.edu/}} (\bibinfo {year} {2026})\BibitemShut {NoStop}%
\bibitem [{\citenamefont {{European Gravitational Observatory}}(2026)}]{EGOwebsite}%
  \BibitemOpen
  \bibfield  {author} {\bibinfo {author} {\bibnamefont {{European Gravitational Observatory}}},\ }\href@noop {} {\bibinfo {title} {{EGO} -- european gravitational observatory}},\ \bibinfo {howpublished} {\url{https://www.ego-gw.it/}} (\bibinfo {year} {2026})\BibitemShut {NoStop}%
\bibitem [{\citenamefont {Abadie}\ \emph {et~al.}(2010)\citenamefont {Abadie} \emph {et~al.}}]{Abadie_2010}%
  \BibitemOpen
  \bibfield  {author} {\bibinfo {author} {\bibfnamefont {J.}~\bibnamefont {Abadie}} \emph {et~al.} (\bibinfo {collaboration} {LIGO Scientific and Virgo}),\ }\bibfield  {title} {\bibinfo {title} {Predictions for the rates of compact binary coalescences observable by ground-based gravitational-wave detectors},\ }\href {https://doi.org/10.1088/0264-9381/27/17/173001} {\bibfield  {journal} {\bibinfo  {journal} {Class. Quantum Grav.}\ }\textbf {\bibinfo {volume} {27}},\ \bibinfo {pages} {173001} (\bibinfo {year} {2010})}\BibitemShut {NoStop}%
\bibitem [{\citenamefont {Somiya}(2012)}]{Somiya_2012}%
  \BibitemOpen
  \bibfield  {author} {\bibinfo {author} {\bibfnamefont {K.}~\bibnamefont {Somiya}} (\bibinfo {collaboration} {KAGRA}),\ }\bibfield  {title} {\bibinfo {title} {Detector configuration of {KAGRA}---the {Japanese} cryogenic gravitational-wave detector},\ }\href {https://doi.org/10.1088/0264-9381/29/12/124007} {\bibfield  {journal} {\bibinfo  {journal} {Class. Quantum Grav.}\ }\textbf {\bibinfo {volume} {29}},\ \bibinfo {pages} {124007} (\bibinfo {year} {2012})}\BibitemShut {NoStop}%
\bibitem [{\citenamefont {Ivanova}\ \emph {et~al.}(2013)\citenamefont {Ivanova} \emph {et~al.}}]{Ivanova:2012vx}%
  \BibitemOpen
  \bibfield  {author} {\bibinfo {author} {\bibfnamefont {N.}~\bibnamefont {Ivanova}} \emph {et~al.},\ }\bibfield  {title} {\bibinfo {title} {{Common Envelope Evolution: Where we stand and how we can move forward}},\ }\href {https://doi.org/10.1007/s00159-013-0059-2} {\bibfield  {journal} {\bibinfo  {journal} {Astron. Astrophys. Rev.}\ }\textbf {\bibinfo {volume} {21}},\ \bibinfo {pages} {59} (\bibinfo {year} {2013})},\ \Eprint {https://arxiv.org/abs/1209.4302} {arXiv:1209.4302 [astro-ph.HE]} \BibitemShut {NoStop}%
\bibitem [{\citenamefont {Clausen}\ \emph {et~al.}(2013)\citenamefont {Clausen}, \citenamefont {Sigurdsson},\ and\ \citenamefont {Chernoff}}]{Clausen:2012zu}%
  \BibitemOpen
  \bibfield  {author} {\bibinfo {author} {\bibfnamefont {D.}~\bibnamefont {Clausen}}, \bibinfo {author} {\bibfnamefont {S.}~\bibnamefont {Sigurdsson}},\ and\ \bibinfo {author} {\bibfnamefont {D.~F.}\ \bibnamefont {Chernoff}},\ }\bibfield  {title} {\bibinfo {title} {{Black Hole-Neutron Star Mergers in Globular Clusters}},\ }\href {https://doi.org/10.1093/mnras/sts295} {\bibfield  {journal} {\bibinfo  {journal} {Mon. Not. Roy. Astron. Soc.}\ }\textbf {\bibinfo {volume} {428}},\ \bibinfo {pages} {3618} (\bibinfo {year} {2013})},\ \Eprint {https://arxiv.org/abs/1210.8153} {arXiv:1210.8153 [astro-ph.HE]} \BibitemShut {NoStop}%
\bibitem [{\citenamefont {Ye}\ \emph {et~al.}(2020)\citenamefont {Ye}, \citenamefont {Fong}, \citenamefont {Kremer}, \citenamefont {Rodriguez}, \citenamefont {Chatterjee}, \citenamefont {Fragione},\ and\ \citenamefont {Rasio}}]{Ye:2019xvf}%
  \BibitemOpen
  \bibfield  {author} {\bibinfo {author} {\bibfnamefont {C.~S.}\ \bibnamefont {Ye}}, \bibinfo {author} {\bibfnamefont {W.-f.}\ \bibnamefont {Fong}}, \bibinfo {author} {\bibfnamefont {K.}~\bibnamefont {Kremer}}, \bibinfo {author} {\bibfnamefont {C.~L.}\ \bibnamefont {Rodriguez}}, \bibinfo {author} {\bibfnamefont {S.}~\bibnamefont {Chatterjee}}, \bibinfo {author} {\bibfnamefont {G.}~\bibnamefont {Fragione}},\ and\ \bibinfo {author} {\bibfnamefont {F.~A.}\ \bibnamefont {Rasio}},\ }\bibfield  {title} {\bibinfo {title} {{On the Rate of Neutron Star Binary Mergers from Globular Clusters}},\ }\href {https://doi.org/10.3847/2041-8213/ab5dc5} {\bibfield  {journal} {\bibinfo  {journal} {Astrophys. J. Lett.}\ }\textbf {\bibinfo {volume} {888}},\ \bibinfo {pages} {L10} (\bibinfo {year} {2020})},\ \Eprint {https://arxiv.org/abs/1910.10740} {arXiv:1910.10740 [astro-ph.HE]} \BibitemShut {NoStop}%
\bibitem [{\citenamefont {Sedda}(2020)}]{Sedda:2020wzl}%
  \BibitemOpen
  \bibfield  {author} {\bibinfo {author} {\bibfnamefont {M.~A.}\ \bibnamefont {Sedda}},\ }\bibfield  {title} {\bibinfo {title} {{Dissecting the properties of neutron star - black hole mergers originating in dense star clusters}},\ }\href {https://doi.org/10.1038/s42005-020-0310-x} {\bibfield  {journal} {\bibinfo  {journal} {Commun. Phys.}\ }\textbf {\bibinfo {volume} {3}},\ \bibinfo {pages} {43} (\bibinfo {year} {2020})},\ \Eprint {https://arxiv.org/abs/2003.02279} {arXiv:2003.02279 [astro-ph.GA]} \BibitemShut {NoStop}%
\bibitem [{\citenamefont {Foucart}(2020)}]{Foucart:2020ats}%
  \BibitemOpen
  \bibfield  {author} {\bibinfo {author} {\bibfnamefont {F.}~\bibnamefont {Foucart}},\ }\bibfield  {title} {\bibinfo {title} {{A brief overview of black hole-neutron star mergers}},\ }\href {https://doi.org/10.3389/fspas.2020.00046} {\bibfield  {journal} {\bibinfo  {journal} {Front. Astron. Space Sci.}\ }\textbf {\bibinfo {volume} {7}},\ \bibinfo {pages} {46} (\bibinfo {year} {2020})},\ \Eprint {https://arxiv.org/abs/2006.10570} {arXiv:2006.10570 [astro-ph.HE]} \BibitemShut {NoStop}%
\bibitem [{\citenamefont {Metzger}(2020)}]{Metzger:2019zeh}%
  \BibitemOpen
  \bibfield  {author} {\bibinfo {author} {\bibfnamefont {B.~D.}\ \bibnamefont {Metzger}},\ }\bibfield  {title} {\bibinfo {title} {{Kilonovae}},\ }\href {https://doi.org/10.1007/s41114-019-0024-0} {\bibfield  {journal} {\bibinfo  {journal} {Living Rev. Rel.}\ }\textbf {\bibinfo {volume} {23}},\ \bibinfo {pages} {1} (\bibinfo {year} {2020})},\ \Eprint {https://arxiv.org/abs/1910.01617} {arXiv:1910.01617 [astro-ph.HE]} \BibitemShut {NoStop}%
\bibitem [{\citenamefont {{Paczynski}}(1986)}]{sgrb1}%
  \BibitemOpen
  \bibfield  {author} {\bibinfo {author} {\bibfnamefont {B.}~\bibnamefont {{Paczynski}}},\ }\bibfield  {title} {\bibinfo {title} {{Gamma-ray bursters at cosmological distances}},\ }\href {https://doi.org/10.1086/184740} {\bibfield  {journal} {\bibinfo  {journal} {Astrophys. J. Lett.}\ }\textbf {\bibinfo {volume} {308}},\ \bibinfo {pages} {L43} (\bibinfo {year} {1986})}\BibitemShut {NoStop}%
\bibitem [{\citenamefont {Eichler}\ \emph {et~al.}(1989)\citenamefont {Eichler}, \citenamefont {Livio}, \citenamefont {Piran},\ and\ \citenamefont {Schramm}}]{sgrb2}%
  \BibitemOpen
  \bibfield  {author} {\bibinfo {author} {\bibfnamefont {D.}~\bibnamefont {Eichler}}, \bibinfo {author} {\bibfnamefont {M.}~\bibnamefont {Livio}}, \bibinfo {author} {\bibfnamefont {T.}~\bibnamefont {Piran}},\ and\ \bibinfo {author} {\bibfnamefont {D.~N.}\ \bibnamefont {Schramm}},\ }\bibfield  {title} {\bibinfo {title} {Nucleosynthesis, neutrino bursts and $\gamma$-rays from coalescing neutron stars},\ }\href {https://doi.org/10.1038/340126a0} {\bibfield  {journal} {\bibinfo  {journal} {Nature}\ }\textbf {\bibinfo {volume} {340}},\ \bibinfo {pages} {126} (\bibinfo {year} {1989})}\BibitemShut {NoStop}%
\bibitem [{\citenamefont {Berger}(2014)}]{Berger:2013jza}%
  \BibitemOpen
  \bibfield  {author} {\bibinfo {author} {\bibfnamefont {E.}~\bibnamefont {Berger}},\ }\bibfield  {title} {\bibinfo {title} {{Short-Duration Gamma-Ray Bursts}},\ }\href {https://doi.org/10.1146/annurev-astro-081913-035926} {\bibfield  {journal} {\bibinfo  {journal} {Ann. Rev. Astron. Astrophys.}\ }\textbf {\bibinfo {volume} {52}},\ \bibinfo {pages} {43} (\bibinfo {year} {2014})},\ \Eprint {https://arxiv.org/abs/1311.2603} {arXiv:1311.2603 [astro-ph.HE]} \BibitemShut {NoStop}%
\bibitem [{\citenamefont {Kyutoku}\ \emph {et~al.}(2021)\citenamefont {Kyutoku}, \citenamefont {Shibata},\ and\ \citenamefont {Taniguchi}}]{Kyutoku:2021icp}%
  \BibitemOpen
  \bibfield  {author} {\bibinfo {author} {\bibfnamefont {K.}~\bibnamefont {Kyutoku}}, \bibinfo {author} {\bibfnamefont {M.}~\bibnamefont {Shibata}},\ and\ \bibinfo {author} {\bibfnamefont {K.}~\bibnamefont {Taniguchi}},\ }\bibfield  {title} {\bibinfo {title} {{Coalescence of black hole{\textendash}neutron star binaries}},\ }\href {https://doi.org/10.1007/s41114-021-00033-4} {\bibfield  {journal} {\bibinfo  {journal} {Living Rev. Rel.}\ }\textbf {\bibinfo {volume} {24}},\ \bibinfo {pages} {5} (\bibinfo {year} {2021})},\ \Eprint {https://arxiv.org/abs/2110.06218} {arXiv:2110.06218 [astro-ph.HE]} \BibitemShut {NoStop}%
\bibitem [{\citenamefont {Matur}\ \emph {et~al.}(2024)\citenamefont {Matur}, \citenamefont {Hawke},\ and\ \citenamefont {Andersson}}]{matur2024signatureslowmassblack}%
  \BibitemOpen
  \bibfield  {author} {\bibinfo {author} {\bibfnamefont {R.}~\bibnamefont {Matur}}, \bibinfo {author} {\bibfnamefont {I.}~\bibnamefont {Hawke}},\ and\ \bibinfo {author} {\bibfnamefont {N.}~\bibnamefont {Andersson}},\ }\href {https://arxiv.org/abs/2407.18045} {\bibinfo {title} {Signatures of low mass black hole-neutron star mergers}} (\bibinfo {year} {2024}),\ \Eprint {https://arxiv.org/abs/2407.18045} {arXiv:2407.18045 [astro-ph.HE]} \BibitemShut {NoStop}%
\bibitem [{\citenamefont {Shibata}\ and\ \citenamefont {Ury{\=u}}(2006)}]{ShibataUryu2006}%
  \BibitemOpen
  \bibfield  {author} {\bibinfo {author} {\bibfnamefont {M.}~\bibnamefont {Shibata}}\ and\ \bibinfo {author} {\bibfnamefont {K.}~\bibnamefont {Ury{\=u}}},\ }\bibfield  {title} {\bibinfo {title} {{Merger of black hole-neutron star binaries: nonspinning black hole case}},\ }\href {https://doi.org/10.1103/PhysRevD.74.121503} {\bibfield  {journal} {\bibinfo  {journal} {Phys. Rev. D}\ }\textbf {\bibinfo {volume} {74}},\ \bibinfo {pages} {121503} (\bibinfo {year} {2006})},\ \Eprint {https://arxiv.org/abs/gr-qc/0612142} {arXiv:gr-qc/0612142} \BibitemShut {NoStop}%
\bibitem [{\citenamefont {Shibata}\ and\ \citenamefont {Uryu}(2007)}]{Shibata:2006bs}%
  \BibitemOpen
  \bibfield  {author} {\bibinfo {author} {\bibfnamefont {M.}~\bibnamefont {Shibata}}\ and\ \bibinfo {author} {\bibfnamefont {K.}~\bibnamefont {Uryu}},\ }\bibfield  {title} {\bibinfo {title} {{Merger of black hole-neutron star binaries in full general relativity}},\ }\href {https://doi.org/10.1088/0264-9381/24/12/S09} {\bibfield  {journal} {\bibinfo  {journal} {Class. Quant. Grav.}\ }\textbf {\bibinfo {volume} {24}},\ \bibinfo {pages} {S125} (\bibinfo {year} {2007})},\ \Eprint {https://arxiv.org/abs/astro-ph/0611522} {arXiv:astro-ph/0611522} \BibitemShut {NoStop}%
\bibitem [{\citenamefont {Etienne}\ \emph {et~al.}(2008)\citenamefont {Etienne}, \citenamefont {Faber}, \citenamefont {Liu}, \citenamefont {Shapiro}, \citenamefont {Taniguchi},\ and\ \citenamefont {Baumgarte}}]{Etienne2008}%
  \BibitemOpen
  \bibfield  {author} {\bibinfo {author} {\bibfnamefont {Z.~B.}\ \bibnamefont {Etienne}}, \bibinfo {author} {\bibfnamefont {J.~A.}\ \bibnamefont {Faber}}, \bibinfo {author} {\bibfnamefont {Y.~T.}\ \bibnamefont {Liu}}, \bibinfo {author} {\bibfnamefont {S.~L.}\ \bibnamefont {Shapiro}}, \bibinfo {author} {\bibfnamefont {K.}~\bibnamefont {Taniguchi}},\ and\ \bibinfo {author} {\bibfnamefont {T.~W.}\ \bibnamefont {Baumgarte}},\ }\bibfield  {title} {\bibinfo {title} {{Fully general relativistic simulations of black hole-neutron star mergers}},\ }\href {https://doi.org/10.1103/PhysRevD.77.084002} {\bibfield  {journal} {\bibinfo  {journal} {Phys. Rev. D}\ }\textbf {\bibinfo {volume} {77}},\ \bibinfo {pages} {084002} (\bibinfo {year} {2008})},\ \Eprint {https://arxiv.org/abs/0712.2460} {arXiv:0712.2460 [astro-ph]} \BibitemShut {NoStop}%
\bibitem [{\citenamefont {Yamamoto}\ \emph {et~al.}(2008)\citenamefont {Yamamoto}, \citenamefont {Shibata},\ and\ \citenamefont {Taniguchi}}]{PhysRevD.78.064054}%
  \BibitemOpen
  \bibfield  {author} {\bibinfo {author} {\bibfnamefont {T.}~\bibnamefont {Yamamoto}}, \bibinfo {author} {\bibfnamefont {M.}~\bibnamefont {Shibata}},\ and\ \bibinfo {author} {\bibfnamefont {K.}~\bibnamefont {Taniguchi}},\ }\bibfield  {title} {\bibinfo {title} {Simulating coalescing compact binaries by a new code (sacra)},\ }\href {https://doi.org/10.1103/PhysRevD.78.064054} {\bibfield  {journal} {\bibinfo  {journal} {Phys. Rev. D}\ }\textbf {\bibinfo {volume} {78}},\ \bibinfo {pages} {064054} (\bibinfo {year} {2008})}\BibitemShut {NoStop}%
\bibitem [{\citenamefont {Foucart}(2012)}]{Foucart2012}%
  \BibitemOpen
  \bibfield  {author} {\bibinfo {author} {\bibfnamefont {F.}~\bibnamefont {Foucart}},\ }\bibfield  {title} {\bibinfo {title} {{Black-hole-neutron-star mergers: Disk mass predictions}},\ }\href {https://doi.org/10.1103/PhysRevD.86.124007} {\bibfield  {journal} {\bibinfo  {journal} {Phys. Rev. D}\ }\textbf {\bibinfo {volume} {86}},\ \bibinfo {pages} {124007} (\bibinfo {year} {2012})},\ \Eprint {https://arxiv.org/abs/1207.6304} {arXiv:1207.6304 [astro-ph.HE]} \BibitemShut {NoStop}%
\bibitem [{\citenamefont {Etienne}\ \emph {et~al.}(2015)\citenamefont {Etienne}, \citenamefont {Paschalidis}, \citenamefont {Haas}, \citenamefont {M{\"o}sta},\ and\ \citenamefont {Shapiro}}]{Etienne2015}%
  \BibitemOpen
  \bibfield  {author} {\bibinfo {author} {\bibfnamefont {Z.~B.}\ \bibnamefont {Etienne}}, \bibinfo {author} {\bibfnamefont {V.}~\bibnamefont {Paschalidis}}, \bibinfo {author} {\bibfnamefont {R.}~\bibnamefont {Haas}}, \bibinfo {author} {\bibfnamefont {P.}~\bibnamefont {M{\"o}sta}},\ and\ \bibinfo {author} {\bibfnamefont {S.~L.}\ \bibnamefont {Shapiro}},\ }\bibfield  {title} {\bibinfo {title} {{IllinoisGRMHD}: An open-source, user-friendly {GRMHD} code for dynamical spacetimes},\ }\href {https://doi.org/10.1088/0264-9381/32/17/175009} {\bibfield  {journal} {\bibinfo  {journal} {Class. Quantum Grav.}\ }\textbf {\bibinfo {volume} {32}},\ \bibinfo {pages} {175009} (\bibinfo {year} {2015})},\ \Eprint {https://arxiv.org/abs/1501.07276} {arXiv:1501.07276} \BibitemShut {NoStop}%
\bibitem [{\citenamefont {{Lattimer}}\ and\ \citenamefont {{Schramm}}(1974)}]{tidal_radius}%
  \BibitemOpen
  \bibfield  {author} {\bibinfo {author} {\bibfnamefont {J.~M.}\ \bibnamefont {{Lattimer}}}\ and\ \bibinfo {author} {\bibfnamefont {D.~N.}\ \bibnamefont {{Schramm}}},\ }\bibfield  {title} {\bibinfo {title} {{Black-Hole-Neutron-Star Collisions}},\ }\href {https://doi.org/10.1086/181612} {\bibfield  {journal} {\bibinfo  {journal} {Astrophys. J. Lett.}\ }\textbf {\bibinfo {volume} {192}},\ \bibinfo {pages} {L145} (\bibinfo {year} {1974})}\BibitemShut {NoStop}%
\bibitem [{\citenamefont {Bardeen}\ \emph {et~al.}(1972)\citenamefont {Bardeen}, \citenamefont {Press},\ and\ \citenamefont {Teukolsky}}]{Bardeen1972}%
  \BibitemOpen
  \bibfield  {author} {\bibinfo {author} {\bibfnamefont {J.~M.}\ \bibnamefont {Bardeen}}, \bibinfo {author} {\bibfnamefont {W.~H.}\ \bibnamefont {Press}},\ and\ \bibinfo {author} {\bibfnamefont {S.~A.}\ \bibnamefont {Teukolsky}},\ }\bibfield  {title} {\bibinfo {title} {Rotating black holes: Locally nonrotating frames, energy extraction, and scalar synchrotron radiation},\ }\href {https://doi.org/10.1086/151796} {\bibfield  {journal} {\bibinfo  {journal} {Astrophys. J.}\ }\textbf {\bibinfo {volume} {178}},\ \bibinfo {pages} {347} (\bibinfo {year} {1972})}\BibitemShut {NoStop}%
\bibitem [{\citenamefont {Kawaguchi}\ \emph {et~al.}(2016)\citenamefont {Kawaguchi}, \citenamefont {Kyutoku}, \citenamefont {Shibata},\ and\ \citenamefont {Tanaka}}]{Kawaguchi:2016ana}%
  \BibitemOpen
  \bibfield  {author} {\bibinfo {author} {\bibfnamefont {K.}~\bibnamefont {Kawaguchi}}, \bibinfo {author} {\bibfnamefont {K.}~\bibnamefont {Kyutoku}}, \bibinfo {author} {\bibfnamefont {M.}~\bibnamefont {Shibata}},\ and\ \bibinfo {author} {\bibfnamefont {M.}~\bibnamefont {Tanaka}},\ }\bibfield  {title} {\bibinfo {title} {{Models of Kilonova/macronova Emission From Black Hole{\textendash}neutron Star Mergers}},\ }\href {https://doi.org/10.3847/0004-637X/825/1/52} {\bibfield  {journal} {\bibinfo  {journal} {Astrophys. J.}\ }\textbf {\bibinfo {volume} {825}},\ \bibinfo {pages} {52} (\bibinfo {year} {2016})},\ \Eprint {https://arxiv.org/abs/1601.07711} {arXiv:1601.07711 [astro-ph.HE]} \BibitemShut {NoStop}%
\bibitem [{\citenamefont {Pannarale}\ \emph {et~al.}(2011)\citenamefont {Pannarale}, \citenamefont {Tonita},\ and\ \citenamefont {Rezzolla}}]{Panarel}%
  \BibitemOpen
  \bibfield  {author} {\bibinfo {author} {\bibfnamefont {F.}~\bibnamefont {Pannarale}}, \bibinfo {author} {\bibfnamefont {A.}~\bibnamefont {Tonita}},\ and\ \bibinfo {author} {\bibfnamefont {L.}~\bibnamefont {Rezzolla}},\ }\bibfield  {title} {\bibinfo {title} {Black hole--neutron star mergers and short gamma-ray bursts: {A} relativistic toy model to estimate the mass of the torus},\ }\href {https://doi.org/10.1088/0004-637X/727/2/95} {\bibfield  {journal} {\bibinfo  {journal} {Astrophys. J.}\ }\textbf {\bibinfo {volume} {727}},\ \bibinfo {pages} {95} (\bibinfo {year} {2011})},\ \Eprint {https://arxiv.org/abs/1007.4160} {arXiv:1007.4160} \BibitemShut {NoStop}%
\bibitem [{\citenamefont {Foucart}\ \emph {et~al.}(2018)\citenamefont {Foucart}, \citenamefont {Hinderer},\ and\ \citenamefont {Nissanke}}]{Foucart2018}%
  \BibitemOpen
  \bibfield  {author} {\bibinfo {author} {\bibfnamefont {F.}~\bibnamefont {Foucart}}, \bibinfo {author} {\bibfnamefont {T.}~\bibnamefont {Hinderer}},\ and\ \bibinfo {author} {\bibfnamefont {S.}~\bibnamefont {Nissanke}},\ }\bibfield  {title} {\bibinfo {title} {Remnant baryon mass in neutron star-black hole mergers: Predictions for binary neutron star mimickers and rapidly spinning black holes},\ }\href {https://doi.org/10.1103/PhysRevD.98.081501} {\bibfield  {journal} {\bibinfo  {journal} {Phys. Rev. D}\ }\textbf {\bibinfo {volume} {98}},\ \bibinfo {pages} {081501(R)} (\bibinfo {year} {2018})}\BibitemShut {NoStop}%
\bibitem [{\citenamefont {Zappa}\ \emph {et~al.}(2019)\citenamefont {Zappa}, \citenamefont {Bernuzzi}, \citenamefont {Pannarale}, \citenamefont {Mapelli},\ and\ \citenamefont {Giacobbo}}]{Zappa2019}%
  \BibitemOpen
  \bibfield  {author} {\bibinfo {author} {\bibfnamefont {F.}~\bibnamefont {Zappa}}, \bibinfo {author} {\bibfnamefont {S.}~\bibnamefont {Bernuzzi}}, \bibinfo {author} {\bibfnamefont {F.}~\bibnamefont {Pannarale}}, \bibinfo {author} {\bibfnamefont {M.}~\bibnamefont {Mapelli}},\ and\ \bibinfo {author} {\bibfnamefont {N.}~\bibnamefont {Giacobbo}},\ }\bibfield  {title} {\bibinfo {title} {{Black-hole remnants from black-hole--neutron-star mergers}},\ }\href {https://doi.org/10.1103/PhysRevLett.123.041102} {\bibfield  {journal} {\bibinfo  {journal} {Phys. Rev. Lett.}\ }\textbf {\bibinfo {volume} {123}},\ \bibinfo {pages} {041102} (\bibinfo {year} {2019})},\ \Eprint {https://arxiv.org/abs/1903.11622} {arXiv:1903.11622} \BibitemShut {NoStop}%
\bibitem [{\citenamefont {Kr{\"u}ger}\ and\ \citenamefont {Foucart}(2020)}]{Kruger2020}%
  \BibitemOpen
  \bibfield  {author} {\bibinfo {author} {\bibfnamefont {C.~J.}\ \bibnamefont {Kr{\"u}ger}}\ and\ \bibinfo {author} {\bibfnamefont {F.}~\bibnamefont {Foucart}},\ }\bibfield  {title} {\bibinfo {title} {{Estimates for disk and ejecta masses produced in compact binary mergers}},\ }\href {https://doi.org/10.1103/PhysRevD.101.103002} {\bibfield  {journal} {\bibinfo  {journal} {Phys. Rev. D}\ }\textbf {\bibinfo {volume} {101}},\ \bibinfo {pages} {103002} (\bibinfo {year} {2020})},\ \Eprint {https://arxiv.org/abs/2002.07728} {arXiv:2002.07728} \BibitemShut {NoStop}%
\bibitem [{\citenamefont {Marck}(1983)}]{Marck1983}%
  \BibitemOpen
  \bibfield  {author} {\bibinfo {author} {\bibfnamefont {J.-A.}\ \bibnamefont {Marck}},\ }\bibfield  {title} {\bibinfo {title} {Solution to the equations of parallel transport in {Kerr} geometry; tidal tensor},\ }\href {https://doi.org/10.1098/rspa.1983.0021} {\bibfield  {journal} {\bibinfo  {journal} {Proc. R. Soc. Lond. A}\ }\textbf {\bibinfo {volume} {385}},\ \bibinfo {pages} {431} (\bibinfo {year} {1983})}\BibitemShut {NoStop}%
\bibitem [{\citenamefont {Xin}\ and\ \citenamefont {Mummery}(2026)}]{XinMummery2026}%
  \BibitemOpen
  \bibfield  {author} {\bibinfo {author} {\bibfnamefont {W.}~\bibnamefont {Xin}}\ and\ \bibinfo {author} {\bibfnamefont {A.}~\bibnamefont {Mummery}},\ }\bibfield  {title} {\bibinfo {title} {Relativistic tidal tensor: General solutions for stationary axisymmetric spacetimes and the hills mass of naked singularities},\ }\href {https://doi.org/10.1103/xdm3-48nb} {\bibfield  {journal} {\bibinfo  {journal} {Phys. Rev. D}\ }\textbf {\bibinfo {volume} {113}},\ \bibinfo {pages} {044036} (\bibinfo {year} {2026})}\BibitemShut {NoStop}%
\bibitem [{\citenamefont {Fishbone}(1973)}]{Fishbone1973}%
  \BibitemOpen
  \bibfield  {author} {\bibinfo {author} {\bibfnamefont {L.~G.}\ \bibnamefont {Fishbone}},\ }\bibfield  {title} {\bibinfo {title} {The relativistic {R}oche problem. {I}. {E}quilibrium theory for a body in equatorial, circular orbit around a {K}err black hole},\ }\href {https://doi.org/10.1086/152395} {\bibfield  {journal} {\bibinfo  {journal} {Astrophys. J.}\ }\textbf {\bibinfo {volume} {185}},\ \bibinfo {pages} {43} (\bibinfo {year} {1973})}\BibitemShut {NoStop}%
\bibitem [{\citenamefont {Wiggins}\ and\ \citenamefont {Lai}(2000)}]{WigginsLai2000}%
  \BibitemOpen
  \bibfield  {author} {\bibinfo {author} {\bibfnamefont {P.}~\bibnamefont {Wiggins}}\ and\ \bibinfo {author} {\bibfnamefont {D.}~\bibnamefont {Lai}},\ }\bibfield  {title} {\bibinfo {title} {Tidal interaction between a fluid star and a {K}err black hole in circular orbit},\ }\href {https://doi.org/10.1086/308544} {\bibfield  {journal} {\bibinfo  {journal} {Astrophys. J.}\ }\textbf {\bibinfo {volume} {532}},\ \bibinfo {pages} {530} (\bibinfo {year} {2000})},\ \Eprint {https://arxiv.org/abs/astro-ph/9907365} {arXiv:astro-ph/9907365} \BibitemShut {NoStop}%
\bibitem [{\citenamefont {Ishii}\ \emph {et~al.}(2005)\citenamefont {Ishii}, \citenamefont {Shibata},\ and\ \citenamefont {Mino}}]{Ishii2005}%
  \BibitemOpen
  \bibfield  {author} {\bibinfo {author} {\bibfnamefont {M.}~\bibnamefont {Ishii}}, \bibinfo {author} {\bibfnamefont {M.}~\bibnamefont {Shibata}},\ and\ \bibinfo {author} {\bibfnamefont {Y.}~\bibnamefont {Mino}},\ }\bibfield  {title} {\bibinfo {title} {Black hole tidal problem in the {F}ermi normal coordinates},\ }\href {https://doi.org/10.1103/PhysRevD.71.044017} {\bibfield  {journal} {\bibinfo  {journal} {Phys. Rev. D}\ }\textbf {\bibinfo {volume} {71}},\ \bibinfo {pages} {044017} (\bibinfo {year} {2005})},\ \Eprint {https://arxiv.org/abs/gr-qc/0501084} {arXiv:gr-qc/0501084} \BibitemShut {NoStop}%
\bibitem [{\citenamefont {Ferrari}\ \emph {et~al.}(2009)\citenamefont {Ferrari}, \citenamefont {Gualtieri},\ and\ \citenamefont {Pannarale}}]{Ferrari2009}%
  \BibitemOpen
  \bibfield  {author} {\bibinfo {author} {\bibfnamefont {V.}~\bibnamefont {Ferrari}}, \bibinfo {author} {\bibfnamefont {L.}~\bibnamefont {Gualtieri}},\ and\ \bibinfo {author} {\bibfnamefont {F.}~\bibnamefont {Pannarale}},\ }\bibfield  {title} {\bibinfo {title} {A semi-relativistic model for tidal interactions in {BH-NS} coalescing binaries},\ }\href {https://doi.org/10.1088/0264-9381/26/12/125004} {\bibfield  {journal} {\bibinfo  {journal} {Class. Quantum Grav.}\ }\textbf {\bibinfo {volume} {26}},\ \bibinfo {pages} {125004} (\bibinfo {year} {2009})},\ \Eprint {https://arxiv.org/abs/0801.2911} {arXiv:0801.2911} \BibitemShut {NoStop}%
\bibitem [{\citenamefont {Ferrari}\ \emph {et~al.}(2010)\citenamefont {Ferrari}, \citenamefont {Gualtieri},\ and\ \citenamefont {Pannarale}}]{Ferrari2010}%
  \BibitemOpen
  \bibfield  {author} {\bibinfo {author} {\bibfnamefont {V.}~\bibnamefont {Ferrari}}, \bibinfo {author} {\bibfnamefont {L.}~\bibnamefont {Gualtieri}},\ and\ \bibinfo {author} {\bibfnamefont {F.}~\bibnamefont {Pannarale}},\ }\bibfield  {title} {\bibinfo {title} {Neutron star tidal disruption in mixed binaries: {T}he imprint of the equation of state},\ }\href {https://doi.org/10.1103/PhysRevD.81.064026} {\bibfield  {journal} {\bibinfo  {journal} {Phys. Rev. D}\ }\textbf {\bibinfo {volume} {81}},\ \bibinfo {pages} {064026} (\bibinfo {year} {2010})},\ \Eprint {https://arxiv.org/abs/0912.3692} {arXiv:0912.3692} \BibitemShut {NoStop}%
\bibitem [{\citenamefont {Taniguchi}\ \emph {et~al.}(2008)\citenamefont {Taniguchi}, \citenamefont {Baumgarte}, \citenamefont {Faber},\ and\ \citenamefont {Shapiro}}]{Taniguchi2008}%
  \BibitemOpen
  \bibfield  {author} {\bibinfo {author} {\bibfnamefont {K.}~\bibnamefont {Taniguchi}}, \bibinfo {author} {\bibfnamefont {T.~W.}\ \bibnamefont {Baumgarte}}, \bibinfo {author} {\bibfnamefont {J.~A.}\ \bibnamefont {Faber}},\ and\ \bibinfo {author} {\bibfnamefont {S.~L.}\ \bibnamefont {Shapiro}},\ }\bibfield  {title} {\bibinfo {title} {{Relativistic black hole-neutron star binaries in quasiequilibrium: effects of the black hole excision boundary condition}},\ }\href {https://doi.org/10.1103/PhysRevD.77.044003} {\bibfield  {journal} {\bibinfo  {journal} {Phys. Rev. D}\ }\textbf {\bibinfo {volume} {77}},\ \bibinfo {pages} {044003} (\bibinfo {year} {2008})},\ \Eprint {https://arxiv.org/abs/0710.5169} {arXiv:0710.5169} \BibitemShut {NoStop}%
\bibitem [{\citenamefont {Masi}(2007)}]{Masi2007}%
  \BibitemOpen
  \bibfield  {author} {\bibinfo {author} {\bibfnamefont {M.}~\bibnamefont {Masi}},\ }\bibfield  {title} {\bibinfo {title} {On compressive radial tidal forces},\ }\href {https://doi.org/10.1119/1.2366736} {\bibfield  {journal} {\bibinfo  {journal} {American Journal of Physics}\ }\textbf {\bibinfo {volume} {75}},\ \bibinfo {pages} {116} (\bibinfo {year} {2007})}\BibitemShut {NoStop}%
\bibitem [{\citenamefont {Weinberg}(1972)}]{Weinberg1972}%
  \BibitemOpen
  \bibfield  {author} {\bibinfo {author} {\bibfnamefont {S.}~\bibnamefont {Weinberg}},\ }\href@noop {} {\emph {\bibinfo {title} {Gravitation and Cosmology: Principles and Applications of the General Theory of Relativity}}}\ (\bibinfo  {publisher} {John Wiley and Sons},\ \bibinfo {address} {New York},\ \bibinfo {year} {1972})\BibitemShut {NoStop}%
\bibitem [{\citenamefont {Pirani}(2009)}]{Pirani2009}%
  \BibitemOpen
  \bibfield  {author} {\bibinfo {author} {\bibfnamefont {F.~A.~E.}\ \bibnamefont {Pirani}},\ }\bibfield  {title} {\bibinfo {title} {Republication of: On the physical significance of the {Riemann} tensor},\ }\href {https://doi.org/10.1007/s10714-009-0787-9} {\bibfield  {journal} {\bibinfo  {journal} {Gen. Relativ. Gravit.}\ }\textbf {\bibinfo {volume} {41}},\ \bibinfo {pages} {1215} (\bibinfo {year} {2009})}\BibitemShut {NoStop}%
\bibitem [{\citenamefont {Guillochon}\ and\ \citenamefont {Ramirez-Ruiz}(2013)}]{Guillochon_2013}%
  \BibitemOpen
  \bibfield  {author} {\bibinfo {author} {\bibfnamefont {J.}~\bibnamefont {Guillochon}}\ and\ \bibinfo {author} {\bibfnamefont {E.}~\bibnamefont {Ramirez-Ruiz}},\ }\bibfield  {title} {\bibinfo {title} {Hydrodynamical simulations to determine the feeding rate of black holes by the tidal disruption of stars: The importance of the impact parameter and stellar structure},\ }\href {https://doi.org/10.1088/0004-637X/767/1/25} {\bibfield  {journal} {\bibinfo  {journal} {The Astrophysical Journal}\ }\textbf {\bibinfo {volume} {767}},\ \bibinfo {pages} {25} (\bibinfo {year} {2013})}\BibitemShut {NoStop}%
\bibitem [{\citenamefont {Oppenheimer}\ and\ \citenamefont {Volkoff}(1939)}]{TOV1}%
  \BibitemOpen
  \bibfield  {author} {\bibinfo {author} {\bibfnamefont {J.~R.}\ \bibnamefont {Oppenheimer}}\ and\ \bibinfo {author} {\bibfnamefont {G.~M.}\ \bibnamefont {Volkoff}},\ }\bibfield  {title} {\bibinfo {title} {On massive neutron cores},\ }\href {https://doi.org/10.1103/PhysRev.55.374} {\bibfield  {journal} {\bibinfo  {journal} {Phys. Rev.}\ }\textbf {\bibinfo {volume} {55}},\ \bibinfo {pages} {374} (\bibinfo {year} {1939})}\BibitemShut {NoStop}%
\bibitem [{\citenamefont {Tolman}(1939)}]{TOV2}%
  \BibitemOpen
  \bibfield  {author} {\bibinfo {author} {\bibfnamefont {R.~C.}\ \bibnamefont {Tolman}},\ }\bibfield  {title} {\bibinfo {title} {Static solutions of einstein's field equations for spheres of fluid},\ }\href {https://doi.org/10.1103/PhysRev.55.364} {\bibfield  {journal} {\bibinfo  {journal} {Phys. Rev.}\ }\textbf {\bibinfo {volume} {55}},\ \bibinfo {pages} {364} (\bibinfo {year} {1939})}\BibitemShut {NoStop}%
\bibitem [{\citenamefont {Stone}\ \emph {et~al.}(2013)\citenamefont {Stone}, \citenamefont {Loeb},\ and\ \citenamefont {Berger}}]{29}%
  \BibitemOpen
  \bibfield  {author} {\bibinfo {author} {\bibfnamefont {N.}~\bibnamefont {Stone}}, \bibinfo {author} {\bibfnamefont {A.}~\bibnamefont {Loeb}},\ and\ \bibinfo {author} {\bibfnamefont {E.}~\bibnamefont {Berger}},\ }\bibfield  {title} {\bibinfo {title} {Pulsations in short gamma ray bursts from black hole-neutron star mergers},\ }\href {https://doi.org/10.1103/PhysRevD.87.084053} {\bibfield  {journal} {\bibinfo  {journal} {Phys. Rev. D}\ }\textbf {\bibinfo {volume} {87}},\ \bibinfo {pages} {084053} (\bibinfo {year} {2013})}\BibitemShut {NoStop}%
\bibitem [{\citenamefont {Mummery}\ and\ \citenamefont {Balbus}(2020)}]{32}%
  \BibitemOpen
  \bibfield  {author} {\bibinfo {author} {\bibfnamefont {A.}~\bibnamefont {Mummery}}\ and\ \bibinfo {author} {\bibfnamefont {S.~A.}\ \bibnamefont {Balbus}},\ }\bibfield  {title} {\bibinfo {title} {{ASASSN-15lh: a TDE about a maximally rotating $10^9\,M_\odot$ black hole}},\ }\href {https://doi.org/10.1093/mnrasl/slaa105} {\bibfield  {journal} {\bibinfo  {journal} {Monthly Notices of the Royal Astronomical Society: Letters}\ }\textbf {\bibinfo {volume} {497}},\ \bibinfo {pages} {L13} (\bibinfo {year} {2020})}\BibitemShut {NoStop}%
\bibitem [{\citenamefont {Ryu}\ \emph {et~al.}(2020)\citenamefont {Ryu}, \citenamefont {Krolik},\ and\ \citenamefont {Piran}}]{33}%
  \BibitemOpen
  \bibfield  {author} {\bibinfo {author} {\bibfnamefont {T.}~\bibnamefont {Ryu}}, \bibinfo {author} {\bibfnamefont {J.}~\bibnamefont {Krolik}},\ and\ \bibinfo {author} {\bibfnamefont {T.}~\bibnamefont {Piran}},\ }\bibfield  {title} {\bibinfo {title} {Measuring stellar and black hole masses of tidal disruption events},\ }\href {https://doi.org/10.3847/1538-4357/abbf4d} {\bibfield  {journal} {\bibinfo  {journal} {The Astrophysical Journal}\ }\textbf {\bibinfo {volume} {904}},\ \bibinfo {pages} {73} (\bibinfo {year} {2020})}\BibitemShut {NoStop}%
\bibitem [{\citenamefont {Chandrasekhar}(1983)}]{Chandrasekhar1983}%
  \BibitemOpen
  \bibfield  {author} {\bibinfo {author} {\bibfnamefont {S.}~\bibnamefont {Chandrasekhar}},\ }\href@noop {} {\emph {\bibinfo {title} {The Mathematical Theory of Black Holes}}},\ \bibinfo {series} {International Series of Monographs on Physics}, Vol.~\bibinfo {volume} {69}\ (\bibinfo  {publisher} {Oxford University Press},\ \bibinfo {address} {Oxford},\ \bibinfo {year} {1983})\BibitemShut {NoStop}%
\bibitem [{\citenamefont {Carter}(1968)}]{Carter1968}%
  \BibitemOpen
  \bibfield  {author} {\bibinfo {author} {\bibfnamefont {B.}~\bibnamefont {Carter}},\ }\bibfield  {title} {\bibinfo {title} {Global structure of the {Kerr} family of gravitational fields},\ }\href {https://doi.org/10.1103/PhysRev.174.1559} {\bibfield  {journal} {\bibinfo  {journal} {Phys. Rev.}\ }\textbf {\bibinfo {volume} {174}},\ \bibinfo {pages} {1559} (\bibinfo {year} {1968})}\BibitemShut {NoStop}%
\bibitem [{\citenamefont {Abbott}\ \emph {et~al.}(2017)\citenamefont {Abbott} \emph {et~al.}}]{GW170817}%
  \BibitemOpen
  \bibfield  {author} {\bibinfo {author} {\bibfnamefont {B.~P.}\ \bibnamefont {Abbott}} \emph {et~al.} (\bibinfo {collaboration} {LIGO Scientific and Virgo}),\ }\bibfield  {title} {\bibinfo {title} {{GW170817}: Observation of gravitational waves from a binary neutron star inspiral},\ }\href {https://doi.org/10.1103/PhysRevLett.119.161101} {\bibfield  {journal} {\bibinfo  {journal} {Phys. Rev. Lett.}\ }\textbf {\bibinfo {volume} {119}},\ \bibinfo {pages} {161101} (\bibinfo {year} {2017})}\BibitemShut {NoStop}%
\bibitem [{\citenamefont {Abbott}\ \emph {et~al.}(2018)\citenamefont {Abbott} \emph {et~al.}}]{GW170817_EOS}%
  \BibitemOpen
  \bibfield  {author} {\bibinfo {author} {\bibfnamefont {B.~P.}\ \bibnamefont {Abbott}} \emph {et~al.} (\bibinfo {collaboration} {LIGO Scientific and Virgo}),\ }\bibfield  {title} {\bibinfo {title} {{GW170817}: Measurements of neutron star radii and equation of state},\ }\href {https://doi.org/10.1103/PhysRevLett.121.161101} {\bibfield  {journal} {\bibinfo  {journal} {Phys. Rev. Lett.}\ }\textbf {\bibinfo {volume} {121}},\ \bibinfo {pages} {161101} (\bibinfo {year} {2018})},\ \Eprint {https://arxiv.org/abs/1805.11581} {arXiv:1805.11581 [gr-qc]} \BibitemShut {NoStop}%
\bibitem [{\citenamefont {Abbott}\ \emph {et~al.}(2019)\citenamefont {Abbott} \emph {et~al.}}]{GW170817_properties}%
  \BibitemOpen
  \bibfield  {author} {\bibinfo {author} {\bibfnamefont {B.~P.}\ \bibnamefont {Abbott}} \emph {et~al.} (\bibinfo {collaboration} {LIGO Scientific and Virgo}),\ }\bibfield  {title} {\bibinfo {title} {Properties of the binary neutron star merger {GW170817}},\ }\href {https://doi.org/10.1103/PhysRevX.9.011001} {\bibfield  {journal} {\bibinfo  {journal} {Phys. Rev. X}\ }\textbf {\bibinfo {volume} {9}},\ \bibinfo {pages} {011001} (\bibinfo {year} {2019})},\ \Eprint {https://arxiv.org/abs/1805.11579} {arXiv:1805.11579 [gr-qc]} \BibitemShut {NoStop}%
\bibitem [{\citenamefont {Sanson}\ \emph {et~al.}(2026)\citenamefont {Sanson}, \citenamefont {Malik},\ and\ \citenamefont {Provid{\^e}ncia}}]{sanson_hyperonic_2026}%
  \BibitemOpen
  \bibfield  {author} {\bibinfo {author} {\bibfnamefont {P.}~\bibnamefont {Sanson}}, \bibinfo {author} {\bibfnamefont {T.}~\bibnamefont {Malik}},\ and\ \bibinfo {author} {\bibfnamefont {C.}~\bibnamefont {Provid{\^e}ncia}},\ }\href {https://doi.org/10.48550/ARXIV.2606.21617} {\bibinfo {title} {Hyperonic equation of state for neutron stars: {{A}} systematic {{Bayesian}} comparison of density-dependent and non-linear relativistic mean-field models}} (\bibinfo {year} {2026}),\ \Eprint {https://arxiv.org/abs/2606.21617} {arXiv:2606.21617 [nucl-th]} \BibitemShut {NoStop}%
\bibitem [{\citenamefont {Malik}\ \emph {et~al.}(2022)\citenamefont {Malik}, \citenamefont {Ferreira}, \citenamefont {Agrawal},\ and\ \citenamefont {Provid{\^e}ncia}}]{Malik:2022}%
  \BibitemOpen
  \bibfield  {author} {\bibinfo {author} {\bibfnamefont {T.}~\bibnamefont {Malik}}, \bibinfo {author} {\bibfnamefont {M.}~\bibnamefont {Ferreira}}, \bibinfo {author} {\bibfnamefont {B.~K.}\ \bibnamefont {Agrawal}},\ and\ \bibinfo {author} {\bibfnamefont {C.}~\bibnamefont {Provid{\^e}ncia}},\ }\bibfield  {title} {\bibinfo {title} {{Relativistic Description of Dense Matter Equation of State and Compatibility with Neutron Star Observables: A Bayesian Approach}},\ }\href {https://doi.org/10.3847/1538-4357/ac5d3c} {\bibfield  {journal} {\bibinfo  {journal} {Astrophys. J.}\ }\textbf {\bibinfo {volume} {930}},\ \bibinfo {pages} {17} (\bibinfo {year} {2022})},\ \Eprint {https://arxiv.org/abs/2201.12552} {arXiv:2201.12552 [nucl-th]} \BibitemShut {NoStop}%
\bibitem [{\citenamefont {Annala}\ \emph {et~al.}(2020)\citenamefont {Annala}, \citenamefont {Gorda}, \citenamefont {Kurkela}, \citenamefont {N{\"a}ttil{\"a}},\ and\ \citenamefont {Vuorinen}}]{Annala:2019puf}%
  \BibitemOpen
  \bibfield  {author} {\bibinfo {author} {\bibfnamefont {E.}~\bibnamefont {Annala}}, \bibinfo {author} {\bibfnamefont {T.}~\bibnamefont {Gorda}}, \bibinfo {author} {\bibfnamefont {A.}~\bibnamefont {Kurkela}}, \bibinfo {author} {\bibfnamefont {J.}~\bibnamefont {N{\"a}ttil{\"a}}},\ and\ \bibinfo {author} {\bibfnamefont {A.}~\bibnamefont {Vuorinen}},\ }\bibfield  {title} {\bibinfo {title} {{Evidence for quark-matter cores in massive neutron stars}},\ }\href {https://doi.org/10.1038/s41567-020-0914-9} {\bibfield  {journal} {\bibinfo  {journal} {Nature Phys.}\ }\textbf {\bibinfo {volume} {16}},\ \bibinfo {pages} {907} (\bibinfo {year} {2020})},\ \Eprint {https://arxiv.org/abs/1903.09121} {arXiv:1903.09121 [astro-ph.HE]} \BibitemShut {NoStop}%
\bibitem [{\citenamefont {Verma}\ \emph {et~al.}(2025)\citenamefont {Verma}, \citenamefont {Saha}, \citenamefont {Malik},\ and\ \citenamefont {Mallick}}]{Verma:2025dez}%
  \BibitemOpen
  \bibfield  {author} {\bibinfo {author} {\bibfnamefont {A.}~\bibnamefont {Verma}}, \bibinfo {author} {\bibfnamefont {A.~K.}\ \bibnamefont {Saha}}, \bibinfo {author} {\bibfnamefont {T.}~\bibnamefont {Malik}},\ and\ \bibinfo {author} {\bibfnamefont {R.}~\bibnamefont {Mallick}},\ }\bibfield  {title} {\bibinfo {title} {{Probing the Internal Structure of Neutron Stars: A Comparative Analysis of Three Different Classes of Equations of State}},\ }\href {https://doi.org/10.3847/1538-4357/ade9a2} {\bibfield  {journal} {\bibinfo  {journal} {Astrophys. J.}\ }\textbf {\bibinfo {volume} {988}},\ \bibinfo {pages} {258} (\bibinfo {year} {2025})},\ \Eprint {https://arxiv.org/abs/2503.16888} {arXiv:2503.16888 [astro-ph.HE]} \BibitemShut {NoStop}%
\bibitem [{\citenamefont {Saha}(2026)}]{asim_kumar_saha_2026}%
  \BibitemOpen
  \bibfield  {author} {\bibinfo {author} {\bibfnamefont {A.~K.}\ \bibnamefont {Saha}},\ }\href {https://doi.org/10.5281/zenodo.20502449} {\bibinfo {title} {Are nicer and gw170817 constraints suggesting a compactified scenario for neutron stars?}} (\bibinfo {year} {2026})\BibitemShut {NoStop}%
\bibitem [{\citenamefont {Kyutoku}\ \emph {et~al.}(2015)\citenamefont {Kyutoku}, \citenamefont {Ioka}, \citenamefont {Okawa}, \citenamefont {Shibata},\ and\ \citenamefont {Taniguchi}}]{NR1}%
  \BibitemOpen
  \bibfield  {author} {\bibinfo {author} {\bibfnamefont {K.}~\bibnamefont {Kyutoku}}, \bibinfo {author} {\bibfnamefont {K.}~\bibnamefont {Ioka}}, \bibinfo {author} {\bibfnamefont {H.}~\bibnamefont {Okawa}}, \bibinfo {author} {\bibfnamefont {M.}~\bibnamefont {Shibata}},\ and\ \bibinfo {author} {\bibfnamefont {K.}~\bibnamefont {Taniguchi}},\ }\bibfield  {title} {\bibinfo {title} {Dynamical mass ejection from black hole-neutron star binaries},\ }\href {https://doi.org/10.1103/PhysRevD.92.044028} {\bibfield  {journal} {\bibinfo  {journal} {Phys. Rev. D}\ }\textbf {\bibinfo {volume} {92}},\ \bibinfo {pages} {044028} (\bibinfo {year} {2015})}\BibitemShut {NoStop}%
\bibitem [{\citenamefont {Foucart}\ \emph {et~al.}(2014)\citenamefont {Foucart}, \citenamefont {Deaton}, \citenamefont {Duez}, \citenamefont {O'Connor}, \citenamefont {Ott}, \citenamefont {Haas}, \citenamefont {Kidder}, \citenamefont {Pfeiffer}, \citenamefont {Scheel},\ and\ \citenamefont {Szilagyi}}]{NR2}%
  \BibitemOpen
  \bibfield  {author} {\bibinfo {author} {\bibfnamefont {F.}~\bibnamefont {Foucart}}, \bibinfo {author} {\bibfnamefont {M.~B.}\ \bibnamefont {Deaton}}, \bibinfo {author} {\bibfnamefont {M.~D.}\ \bibnamefont {Duez}}, \bibinfo {author} {\bibfnamefont {E.}~\bibnamefont {O'Connor}}, \bibinfo {author} {\bibfnamefont {C.~D.}\ \bibnamefont {Ott}}, \bibinfo {author} {\bibfnamefont {R.}~\bibnamefont {Haas}}, \bibinfo {author} {\bibfnamefont {L.~E.}\ \bibnamefont {Kidder}}, \bibinfo {author} {\bibfnamefont {H.~P.}\ \bibnamefont {Pfeiffer}}, \bibinfo {author} {\bibfnamefont {M.~A.}\ \bibnamefont {Scheel}},\ and\ \bibinfo {author} {\bibfnamefont {B.}~\bibnamefont {Szilagyi}},\ }\bibfield  {title} {\bibinfo {title} {Neutron star-black hole mergers with a nuclear equation of state and neutrino cooling: Dependence in the binary parameters},\ }\href {https://doi.org/10.1103/PhysRevD.90.024026} {\bibfield  {journal} {\bibinfo  {journal} {Phys. Rev. D}\ }\textbf {\bibinfo {volume} {90}},\ \bibinfo {pages} {024026} (\bibinfo
  {year} {2014})}\BibitemShut {NoStop}%
\bibitem [{\citenamefont {Brege}\ \emph {et~al.}(2018)\citenamefont {Brege}, \citenamefont {Duez}, \citenamefont {Foucart}, \citenamefont {Deaton}, \citenamefont {Caro}, \citenamefont {Hemberger}, \citenamefont {Kidder}, \citenamefont {O'Connor}, \citenamefont {Pfeiffer},\ and\ \citenamefont {Scheel}}]{NR3}%
  \BibitemOpen
  \bibfield  {author} {\bibinfo {author} {\bibfnamefont {W.}~\bibnamefont {Brege}}, \bibinfo {author} {\bibfnamefont {M.~D.}\ \bibnamefont {Duez}}, \bibinfo {author} {\bibfnamefont {F.}~\bibnamefont {Foucart}}, \bibinfo {author} {\bibfnamefont {M.~B.}\ \bibnamefont {Deaton}}, \bibinfo {author} {\bibfnamefont {J.}~\bibnamefont {Caro}}, \bibinfo {author} {\bibfnamefont {D.~A.}\ \bibnamefont {Hemberger}}, \bibinfo {author} {\bibfnamefont {L.~E.}\ \bibnamefont {Kidder}}, \bibinfo {author} {\bibfnamefont {E.}~\bibnamefont {O'Connor}}, \bibinfo {author} {\bibfnamefont {H.~P.}\ \bibnamefont {Pfeiffer}},\ and\ \bibinfo {author} {\bibfnamefont {M.~A.}\ \bibnamefont {Scheel}},\ }\bibfield  {title} {\bibinfo {title} {Black hole-neutron star mergers using a survey of finite-temperature equations of state},\ }\href {https://doi.org/10.1103/PhysRevD.98.063009} {\bibfield  {journal} {\bibinfo  {journal} {Phys. Rev. D}\ }\textbf {\bibinfo {volume} {98}},\ \bibinfo {pages} {063009} (\bibinfo {year} {2018})}\BibitemShut
  {NoStop}%
\bibitem [{\citenamefont {Lovelace}\ \emph {et~al.}(2013)\citenamefont {Lovelace}, \citenamefont {Duez}, \citenamefont {Foucart}, \citenamefont {Kidder}, \citenamefont {Pfeiffer}, \citenamefont {Scheel},\ and\ \citenamefont {Szil{\'a}gyi}}]{NR4}%
  \BibitemOpen
  \bibfield  {author} {\bibinfo {author} {\bibfnamefont {G.}~\bibnamefont {Lovelace}}, \bibinfo {author} {\bibfnamefont {M.~D.}\ \bibnamefont {Duez}}, \bibinfo {author} {\bibfnamefont {F.}~\bibnamefont {Foucart}}, \bibinfo {author} {\bibfnamefont {L.~E.}\ \bibnamefont {Kidder}}, \bibinfo {author} {\bibfnamefont {H.~P.}\ \bibnamefont {Pfeiffer}}, \bibinfo {author} {\bibfnamefont {M.~A.}\ \bibnamefont {Scheel}},\ and\ \bibinfo {author} {\bibfnamefont {B.}~\bibnamefont {Szil{\'a}gyi}},\ }\bibfield  {title} {\bibinfo {title} {Massive disc formation in the tidal disruption of a neutron star by a nearly extremal black hole},\ }\href {https://doi.org/10.1088/0264-9381/30/13/135004} {\bibfield  {journal} {\bibinfo  {journal} {Class. Quantum Grav.}\ }\textbf {\bibinfo {volume} {30}},\ \bibinfo {pages} {135004} (\bibinfo {year} {2013})}\BibitemShut {NoStop}%
\bibitem [{\citenamefont {Foucart}\ \emph {et~al.}(2012)\citenamefont {Foucart}, \citenamefont {Duez}, \citenamefont {Kidder}, \citenamefont {Scheel}, \citenamefont {Szilagyi},\ and\ \citenamefont {Teukolsky}}]{NR5}%
  \BibitemOpen
  \bibfield  {author} {\bibinfo {author} {\bibfnamefont {F.}~\bibnamefont {Foucart}}, \bibinfo {author} {\bibfnamefont {M.~D.}\ \bibnamefont {Duez}}, \bibinfo {author} {\bibfnamefont {L.~E.}\ \bibnamefont {Kidder}}, \bibinfo {author} {\bibfnamefont {M.~A.}\ \bibnamefont {Scheel}}, \bibinfo {author} {\bibfnamefont {B.}~\bibnamefont {Szilagyi}},\ and\ \bibinfo {author} {\bibfnamefont {S.~A.}\ \bibnamefont {Teukolsky}},\ }\bibfield  {title} {\bibinfo {title} {Black hole-neutron star mergers for $10\,m_\odot$ black holes},\ }\href {https://doi.org/10.1103/PhysRevD.85.044015} {\bibfield  {journal} {\bibinfo  {journal} {Phys. Rev. D}\ }\textbf {\bibinfo {volume} {85}},\ \bibinfo {pages} {044015} (\bibinfo {year} {2012})}\BibitemShut {NoStop}%
\bibitem [{\citenamefont {Foucart}\ \emph {et~al.}(2011)\citenamefont {Foucart}, \citenamefont {Duez}, \citenamefont {Kidder},\ and\ \citenamefont {Teukolsky}}]{NR6}%
  \BibitemOpen
  \bibfield  {author} {\bibinfo {author} {\bibfnamefont {F.}~\bibnamefont {Foucart}}, \bibinfo {author} {\bibfnamefont {M.~D.}\ \bibnamefont {Duez}}, \bibinfo {author} {\bibfnamefont {L.~E.}\ \bibnamefont {Kidder}},\ and\ \bibinfo {author} {\bibfnamefont {S.~A.}\ \bibnamefont {Teukolsky}},\ }\bibfield  {title} {\bibinfo {title} {Black hole-neutron star mergers: Effects of the orientation of the black hole spin},\ }\href {https://doi.org/10.1103/PhysRevD.83.024005} {\bibfield  {journal} {\bibinfo  {journal} {Phys. Rev. D}\ }\textbf {\bibinfo {volume} {83}},\ \bibinfo {pages} {024005} (\bibinfo {year} {2011})}\BibitemShut {NoStop}%
\bibitem [{\citenamefont {Kyutoku}\ \emph {et~al.}(2011)\citenamefont {Kyutoku}, \citenamefont {Okawa}, \citenamefont {Shibata},\ and\ \citenamefont {Taniguchi}}]{NR7}%
  \BibitemOpen
  \bibfield  {author} {\bibinfo {author} {\bibfnamefont {K.}~\bibnamefont {Kyutoku}}, \bibinfo {author} {\bibfnamefont {H.}~\bibnamefont {Okawa}}, \bibinfo {author} {\bibfnamefont {M.}~\bibnamefont {Shibata}},\ and\ \bibinfo {author} {\bibfnamefont {K.}~\bibnamefont {Taniguchi}},\ }\bibfield  {title} {\bibinfo {title} {Gravitational waves from spinning black hole-neutron star binaries: dependence on black hole spins and on neutron star equations of state},\ }\href {https://doi.org/10.1103/PhysRevD.84.064018} {\bibfield  {journal} {\bibinfo  {journal} {Phys. Rev. D}\ }\textbf {\bibinfo {volume} {84}},\ \bibinfo {pages} {064018} (\bibinfo {year} {2011})}\BibitemShut {NoStop}%
\bibitem [{\citenamefont {Etienne}\ \emph {et~al.}(2009)\citenamefont {Etienne}, \citenamefont {Liu}, \citenamefont {Shapiro},\ and\ \citenamefont {Baumgarte}}]{NR8}%
  \BibitemOpen
  \bibfield  {author} {\bibinfo {author} {\bibfnamefont {Z.~B.}\ \bibnamefont {Etienne}}, \bibinfo {author} {\bibfnamefont {Y.~T.}\ \bibnamefont {Liu}}, \bibinfo {author} {\bibfnamefont {S.~L.}\ \bibnamefont {Shapiro}},\ and\ \bibinfo {author} {\bibfnamefont {T.~W.}\ \bibnamefont {Baumgarte}},\ }\bibfield  {title} {\bibinfo {title} {General relativistic simulations of black-hole--neutron-star mergers: Effects of black-hole spin},\ }\href {https://doi.org/10.1103/PhysRevD.79.044024} {\bibfield  {journal} {\bibinfo  {journal} {Phys. Rev. D}\ }\textbf {\bibinfo {volume} {79}},\ \bibinfo {pages} {044024} (\bibinfo {year} {2009})}\BibitemShut {NoStop}%
\bibitem [{\citenamefont {Foucart}\ \emph {et~al.}(2013)\citenamefont {Foucart}, \citenamefont {Deaton}, \citenamefont {Duez}, \citenamefont {Kidder}, \citenamefont {MacDonald}, \citenamefont {Ott}, \citenamefont {Pfeiffer}, \citenamefont {Scheel}, \citenamefont {Szilagyi},\ and\ \citenamefont {Teukolsky}}]{NR9}%
  \BibitemOpen
  \bibfield  {author} {\bibinfo {author} {\bibfnamefont {F.}~\bibnamefont {Foucart}}, \bibinfo {author} {\bibfnamefont {M.~B.}\ \bibnamefont {Deaton}}, \bibinfo {author} {\bibfnamefont {M.~D.}\ \bibnamefont {Duez}}, \bibinfo {author} {\bibfnamefont {L.~E.}\ \bibnamefont {Kidder}}, \bibinfo {author} {\bibfnamefont {I.}~\bibnamefont {MacDonald}}, \bibinfo {author} {\bibfnamefont {C.~D.}\ \bibnamefont {Ott}}, \bibinfo {author} {\bibfnamefont {H.~P.}\ \bibnamefont {Pfeiffer}}, \bibinfo {author} {\bibfnamefont {M.~A.}\ \bibnamefont {Scheel}}, \bibinfo {author} {\bibfnamefont {B.}~\bibnamefont {Szilagyi}},\ and\ \bibinfo {author} {\bibfnamefont {S.~A.}\ \bibnamefont {Teukolsky}},\ }\bibfield  {title} {\bibinfo {title} {Black-hole--neutron-star mergers at realistic mass ratios: Equation of state and spin orientation effects},\ }\href {https://doi.org/10.1103/PhysRevD.87.084006} {\bibfield  {journal} {\bibinfo  {journal} {Phys. Rev. D}\ }\textbf {\bibinfo {volume} {87}},\ \bibinfo {pages} {084006} (\bibinfo {year}
  {2013})}\BibitemShut {NoStop}%
\bibitem [{\citenamefont {Read}\ \emph {et~al.}(2009)\citenamefont {Read}, \citenamefont {Lackey}, \citenamefont {Owen},\ and\ \citenamefont {Friedman}}]{Read2009}%
  \BibitemOpen
  \bibfield  {author} {\bibinfo {author} {\bibfnamefont {J.~S.}\ \bibnamefont {Read}}, \bibinfo {author} {\bibfnamefont {B.~D.}\ \bibnamefont {Lackey}}, \bibinfo {author} {\bibfnamefont {B.~J.}\ \bibnamefont {Owen}},\ and\ \bibinfo {author} {\bibfnamefont {J.~L.}\ \bibnamefont {Friedman}},\ }\bibfield  {title} {\bibinfo {title} {Constraints on a phenomenologically parametrized neutron-star equation of state},\ }\href {https://doi.org/10.1103/PhysRevD.79.124032} {\bibfield  {journal} {\bibinfo  {journal} {Phys. Rev. D}\ }\textbf {\bibinfo {volume} {79}},\ \bibinfo {pages} {124032} (\bibinfo {year} {2009})},\ \Eprint {https://arxiv.org/abs/0812.2163} {arXiv:0812.2163 [astro-ph]} \BibitemShut {NoStop}%
\bibitem [{\citenamefont {Lattimer}\ and\ \citenamefont {Prakash}(2016)}]{LattimerPrakash2016}%
  \BibitemOpen
  \bibfield  {author} {\bibinfo {author} {\bibfnamefont {J.~M.}\ \bibnamefont {Lattimer}}\ and\ \bibinfo {author} {\bibfnamefont {M.}~\bibnamefont {Prakash}},\ }\bibfield  {title} {\bibinfo {title} {The equation of state of hot, dense matter and neutron stars},\ }\href {https://doi.org/10.1016/j.physrep.2015.12.005} {\bibfield  {journal} {\bibinfo  {journal} {Phys. Rep.}\ }\textbf {\bibinfo {volume} {621}},\ \bibinfo {pages} {127} (\bibinfo {year} {2016})},\ \Eprint {https://arxiv.org/abs/1512.07820} {arXiv:1512.07820 [astro-ph.SR]} \BibitemShut {NoStop}%
\bibitem [{\citenamefont {Antoniadis}\ \emph {et~al.}(2013)\citenamefont {Antoniadis} \emph {et~al.}}]{Antoniadis2013}%
  \BibitemOpen
  \bibfield  {author} {\bibinfo {author} {\bibfnamefont {J.}~\bibnamefont {Antoniadis}} \emph {et~al.},\ }\bibfield  {title} {\bibinfo {title} {A massive pulsar in a compact relativistic binary},\ }\href {https://doi.org/10.1126/science.1233232} {\bibfield  {journal} {\bibinfo  {journal} {Science}\ }\textbf {\bibinfo {volume} {340}},\ \bibinfo {pages} {1233232} (\bibinfo {year} {2013})},\ \Eprint {https://arxiv.org/abs/1304.6875} {arXiv:1304.6875 [astro-ph.HE]} \BibitemShut {NoStop}%
\bibitem [{\citenamefont {Fonseca}\ \emph {et~al.}(2021)\citenamefont {Fonseca} \emph {et~al.}}]{Fonseca2021}%
  \BibitemOpen
  \bibfield  {author} {\bibinfo {author} {\bibfnamefont {E.}~\bibnamefont {Fonseca}} \emph {et~al.},\ }\bibfield  {title} {\bibinfo {title} {Refined mass and geometric measurements of the high-mass {PSR J0740+6620}},\ }\href {https://doi.org/10.3847/2041-8213/ac03b8} {\bibfield  {journal} {\bibinfo  {journal} {Astrophys. J. Lett.}\ }\textbf {\bibinfo {volume} {915}},\ \bibinfo {pages} {L12} (\bibinfo {year} {2021})},\ \Eprint {https://arxiv.org/abs/2104.00880} {arXiv:2104.00880 [astro-ph.HE]} \BibitemShut {NoStop}%
\bibitem [{\citenamefont {Miller}\ \emph {et~al.}(2019)\citenamefont {Miller}, \citenamefont {Lamb}, \citenamefont {Dittmann}, \citenamefont {Bogdanov}, \citenamefont {Arzoumanian}, \citenamefont {Gendreau}, \citenamefont {Guillot}, \citenamefont {Harding}, \citenamefont {Ho}, \citenamefont {Lattimer}, \citenamefont {Ludlam}, \citenamefont {Mahmoodifar}, \citenamefont {Morsink}, \citenamefont {Ray}, \citenamefont {Strohmayer}, \citenamefont {Wood}, \citenamefont {Enoto}, \citenamefont {Foster}, \citenamefont {Okajima}, \citenamefont {Prigozhin},\ and\ \citenamefont {Soong}}]{Miller_2019}%
  \BibitemOpen
  \bibfield  {author} {\bibinfo {author} {\bibfnamefont {M.~C.}\ \bibnamefont {Miller}}, \bibinfo {author} {\bibfnamefont {F.~K.}\ \bibnamefont {Lamb}}, \bibinfo {author} {\bibfnamefont {A.~J.}\ \bibnamefont {Dittmann}}, \bibinfo {author} {\bibfnamefont {S.}~\bibnamefont {Bogdanov}}, \bibinfo {author} {\bibfnamefont {Z.}~\bibnamefont {Arzoumanian}}, \bibinfo {author} {\bibfnamefont {K.~C.}\ \bibnamefont {Gendreau}}, \bibinfo {author} {\bibfnamefont {S.}~\bibnamefont {Guillot}}, \bibinfo {author} {\bibfnamefont {A.~K.}\ \bibnamefont {Harding}}, \bibinfo {author} {\bibfnamefont {W.~C.~G.}\ \bibnamefont {Ho}}, \bibinfo {author} {\bibfnamefont {J.~M.}\ \bibnamefont {Lattimer}}, \bibinfo {author} {\bibfnamefont {R.~M.}\ \bibnamefont {Ludlam}}, \bibinfo {author} {\bibfnamefont {S.}~\bibnamefont {Mahmoodifar}}, \bibinfo {author} {\bibfnamefont {S.~M.}\ \bibnamefont {Morsink}}, \bibinfo {author} {\bibfnamefont {P.~S.}\ \bibnamefont {Ray}}, \bibinfo {author} {\bibfnamefont {T.~E.}\ \bibnamefont {Strohmayer}}, \bibinfo
  {author} {\bibfnamefont {K.~S.}\ \bibnamefont {Wood}}, \bibinfo {author} {\bibfnamefont {T.}~\bibnamefont {Enoto}}, \bibinfo {author} {\bibfnamefont {R.}~\bibnamefont {Foster}}, \bibinfo {author} {\bibfnamefont {T.}~\bibnamefont {Okajima}}, \bibinfo {author} {\bibfnamefont {G.}~\bibnamefont {Prigozhin}},\ and\ \bibinfo {author} {\bibfnamefont {Y.}~\bibnamefont {Soong}},\ }\bibfield  {title} {\bibinfo {title} {Psr j0030+0451 mass and radius from nicer data and implications for the properties of neutron star matter},\ }\href {https://doi.org/10.3847/2041-8213/ab50c5} {\bibfield  {journal} {\bibinfo  {journal} {The Astrophysical Journal Letters}\ }\textbf {\bibinfo {volume} {887}},\ \bibinfo {pages} {L24} (\bibinfo {year} {2019})}\BibitemShut {NoStop}%
\end{thebibliography}%
\end{document}